\documentclass[%
 aps,
 nopacs,
 onecolumn,
 preprintnumbers,
 superscriptaddress,
 amsmath,amssymb,
]{revtex4}

\usepackage{graphicx}
\usepackage{dcolumn}
\usepackage{bm}
\usepackage{hyperref}
\usepackage{color}

\begin{document} 

%
%
%
%
%

\title{Scaling of the energy gap in pattern-hydrogenated graphene}

\author{Roberto Grassi}
\email[E-mail: ]{rgrassi@arces.unibo.it}
\affiliation{School of Electrical \& Computer Engineering, Purdue University, West Lafayette, Indiana 47906, USA}
\affiliation{ARCES and DEIS, University of Bologna, Viale Risorgimento 2, 40136 Bologna, Italy}

\author{Tony Low}
\altaffiliation[Current address: ]{IBM T.J. Watson Research Center, Yorktown Heights, New York 10598, USA}
\affiliation{School of Electrical \& Computer Engineering, Purdue University, West Lafayette, Indiana 47906, USA}

\author{Mark Lundstrom}
\affiliation{School of Electrical \& Computer Engineering, Purdue University, West Lafayette, Indiana 47906, USA}

\date{\today}


\begin{abstract}
Recent experiments show that a substantial energy gap in graphene can be induced via patterned hydrogenation on an iridium substrate. Here, 
we show that the energy gap is roughly proportional to $\sqrt{N_{H}}/N_{C}$ when disorder is accounted for, where $N_H$ and $N_C$ denote concentration of hydrogen and carbon atoms, respectively.
The dispersion relation, obtained through calculation of the momentum-energy resolved density of states, is shown to agree with 
previous angle-resolved photoemission spectroscopy results. Simulations of electronic transport in finite size samples
also reveal a similar transport gap, up to 1~eV within experimentally achievable $\sqrt{N_{H}}/N_{C}$ value.

\end{abstract}

\maketitle 



\textbf{Introduction:} Graphene's novel electronic and physical properties makes it an interesting material for various potential applications \cite{geim07,neto09}. For electronics, the absence of an energy gap limits its applicability. Currently, there are myriad known ways to opening an energy gap in graphene. For instance, the patterning of graphene nanoribbons induces a band gap due to the confinement of carriers along the transverse direction \cite{han07}. Although there are recent proposals in achieving controlled width and smooth edges \cite{wang08}, the large scale fabrication of nanoribbons remains a challenge. The graphene nanomesh, also known as a graphene antidot lattice, is a viable alternative \cite{pedersen08,bai10}. Here, the confinement potential is created by clusters of vacancies, i.e. nanoholes, arranged on a regular superlattice. They are prepared using block copolymer lithography. Two structural parameters, the cluster size and neck width of the superlattice of nanoholes, govern the electronic transport properties observed in experiments \cite{bai10}. A larger energy gap is induced when both these two parameters are reduced.

Another variant of nanomesh, formed by periodic pattern of hydrogen clusters, has recently been observed for graphene grown on an iridium substrate \cite{balog10}. The periodicity is due to the fact that the composite structure of graphene and iridium forms a superlattice, with the hydrogenation occuring preferentially on specific superlattice sites. The resulting structure can be regarded as a variant of the nanomesh since regions of hydrogenated graphene are highly insulating \cite{sofo07,elias09,fiori10}. Nanomesh via patterned hydrogenation is a promising approach since its cluster size and neck width can be much smaller than the lithographically defined case. Indeed, the opening of a substantial energy gap has been revealed by angle-resolved photoemission spectroscopy (ARPES) \cite{balog10}.


Theoretical investigations of the electronic properties of nanomesh/pattern-hydrogenated graphene have been limited so far to band-structure calculations using primitive supercells \cite{pedersen08,petersen11,balog10,yang10}. This approach can treat disorder in the cluster shape only within the same supercell \cite{balog10}. Instead, in this work, we present a modeling study of pattern-hydrogenated graphene that also includes disorder across different supercells. Through calculations of the momentum-energy resolved density of states and its electrical conduction, we study the scaling of the energy gap on the parameters defining the patterned hydrogenation, i.e. cluster size, filling factor and neck width. \\

\textbf{The model:} A simple tight-binding model is employed to describe the composite structure of graphene with adsorbed hydrogen atoms \cite{robinson08,bang10}. Within this model, the basis consists of a $2p_z$ orbital per carbon atom and a $1s$ orbital per hydrogen atom: the parameters describing the carbon-carbon hopping integral ($\gamma = 2.6$ eV), carbon-hydrogen hopping integral ($\gamma_H = 5.72$ eV), and hydrogen onsite energy ($\epsilon_H = 0$ eV) are taken from Ref.~\onlinecite{bang10}. Such a minimal model captures the essential physics of the hydrogenation effect pertinent for our study, that is the removal of the $p_z$ orbital of the hydrogenated carbon atom from the $\pi$ and $\pi^*$ bands \footnote{As explained in Ref.~\onlinecite{robinson08}, the effect of each hydrogen atom at energy $E$ can be recast in a effective self-energy $\Sigma_H = \gamma_H^2/\left( E+i 0^+ -\epsilon_H \right)$ for the attached carbon orbital: since $\Sigma_H \gg 1$ eV for $0 < |E| < 1$ eV, the hydrogenated carbon atom effectively acts as a vacancy in the energy range of interest.}. Since our purpose is to study the intrinsic properties of the nanomesh, we neglect the interaction with the iridium substrate in the TB model. As a consequence, particle-hole symmetry is preserved and the neutrality point remains located at the Dirac point. 

\begin{figure*}[htps]
\includegraphics[scale=0.495]{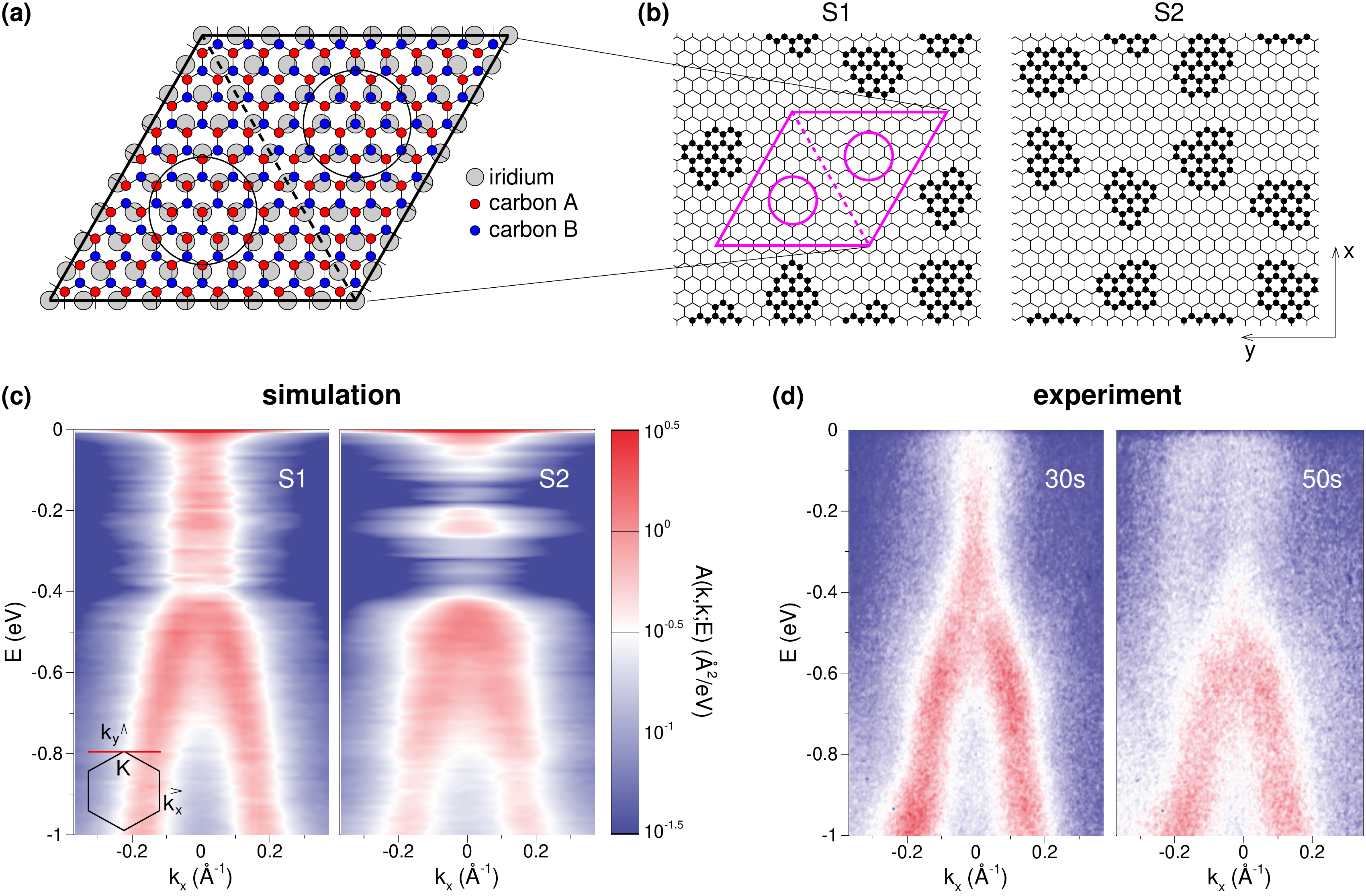}
\caption{\label{fig_structure_arpes}Schematic representation of the atomic structure under study and comparison of the simulated $k$-resolved density of states in energy with the experimental ARPES. (a) Top view of the supercell of graphene on iridium substrate. The two graphene sublattices are indicated with different colors. The supercell is symmetric under reflection across the dashed line, except for the interchange of the two graphene sublattices. The two circles highlight the regions of the supercell where the clusters tend to form. (b) Top view of two hydrogenated samples with different cluster concentration. Hydrogen atoms are represented as black dots on the honeycomb graphene lattice and the iridium substrate is not shown. S1 is obtained with the model parameters $N_w = 4$, $n_c = 0.75$, while S2 with $N_w = 4$, $n_c = 1$. (c) Calculated momentum-energy resolved density of states for two sets of hydrogenated samples that correspond to the cases S1 and S2 shown in (b). 50 samples are considered for each set, the plotted quantity being the average. The inset shows the direction within the graphene Brillouin zone (red line) along which the calculation is performed. (d) Experimental ARPES intensity for different times of exposure to hydrogen (as indicated in the labels), reprinted by permission from Macmillan Publishers Ltd: \href{http://www.nature.com/nmat}{Nature Materials} \textbf{9}, 315, copyright 2010.}
\end{figure*}

Graphene on an iridium substrate forms a superlattice due to the mismatch between their respective lattice constants; 10$\times$10 graphene unit cells are commensurate with 9$\times$9 iridium unit cells \cite{ndiaye06}. The superlattice unit cell is represented in Fig.~\ref{fig_structure_arpes}a. The supercell preserves the symmetry of the graphene unit cell \footnote{It can be divided into two regions that are equivalent to each other under reflection, apart from the exchange of the two graphene sublattices.}, resulting in a honeycomb superlattice. Experimentally \cite{balog10}, it was shown that the hydrogen clusters tend to form in regions indicated by circles in Fig.~\ref{fig_structure_arpes}a, where one graphene sublattice sits directly on top of iridium atoms. A hydrogenation model is developed to reproduce this preferential adsorption (see details in Supp. info). The model takes as input two parameters: a discrete quantity $N_w$ which represents the cluster radius, and the filling factor $n_c$ i.e. ratio between the number of clusters and the number of half-supercells. Two types of disorder are considered: $i$) irregular cluster edges and $ii$) a random filling of the superlattice (if $n_c < 1$). See Fig.~\ref{fig_structure_arpes}b for illustrations. Each cluster is generated by adding hydrogen atoms on top of the carbon atoms belonging to a certain number of shells around the center of the half-supercell, and the edge disorder is introduced by partial hydrogenation of the outer shell. Both $A$ and $B$ sites are hydrogenated within each cluster, contrary to what is expected in the real structure \cite{balog10}. The additional hydrogen atoms play the role of the neglected interactions with the substrate in removing the pseudo dangling bonds \cite{balog10,brako10}.\\


\textbf{Key features in momentum-energy resolved density of states:} In order to study the electronic properties of pattern-hydrogenated graphene as seen in ARPES experiments, the calculation of the momentum-energy resolved density of states is required. This quantity is given, apart from a normalization factor, by the diagonal elements of the spectral function in momentum space, $A(\mathbf{k},\mathbf{k};E)$. While this quantity reduces to the usual band structure for periodic systems, it is a general concept and is valid even for disordered systems. The calculation is performed by first computing the spectral function in real space $A(\mathbf{r},\mathbf{r}';E)$ and then Fourier transforming to get $A(\mathbf{k},\mathbf{k};E)$ \footnote{More precisely, within the TB representation, the continuous position $\mathbf{r}$ is substituted by the pair $(\mathbf{l},q)$, with $\mathbf{l}$ the graphene lattice vector and $q$ the orbital index: the projected spectral function for carbon atoms only is considered ($q=1,2$), the Fourier transform is done with respect to $\mathbf{l}$, and the result is summed over $q$, that is over the two graphene sublattices (an additional factor is used in our plots to normalize the number of states with respect to the graphene unit cell; spin degeneracy is not included).}. The calculation is done using the Green's function formalism with a recursive algorithm for periodic structures. See Supp. info for detailed numerical description and implementation.


In Fig.~\ref{fig_structure_arpes}c, we plot the averaged $A(\mathbf{k},\mathbf{k};E)$ for two ensembles corresponding to the two realizations shown in Fig.~\ref{fig_structure_arpes}b, along a path in $k$-space that includes the $K$ point \footnote{We note the absence of repeating dispersions expected for a periodic superlattice structure. This is due to the effect of disorder, which destroys the strict periodicity.}. The convergence of the result with respect to sample size and ensemble size has been checked, as reported in Supp. info. Only the negative energies are shown, as the conduction bands are symmetrical to the valence ones due to particle-hole symmetry. The corresponding experimental ARPES image \cite{balog10} corresponding to two different hydrogen doses is shown in Fig.~\ref{fig_structure_arpes}d. Several distinctive features are observed in both simulations and experiments. In both cases, the two valence branches intersect at a lower energy than the Dirac point. In addition, the signal of the states lying at the $K$ point between $E=0$ and the intersection energy gets suppressed with increasing hydrogen doping. These features can be interpreted as a band-gap opening. The presence of a flat band at $E=0$ in the simulation results is a well-known effect, due to the imbalance between the two graphene sublattices \cite{pereira08}. The absence of these states in the experimental ARPES could be related to bond relaxation and $sp^3$ hybridization \cite{boukhvalov08}, which are neglected in the simulations.\\

\begin{figure}[t]
\includegraphics[scale=0.5]{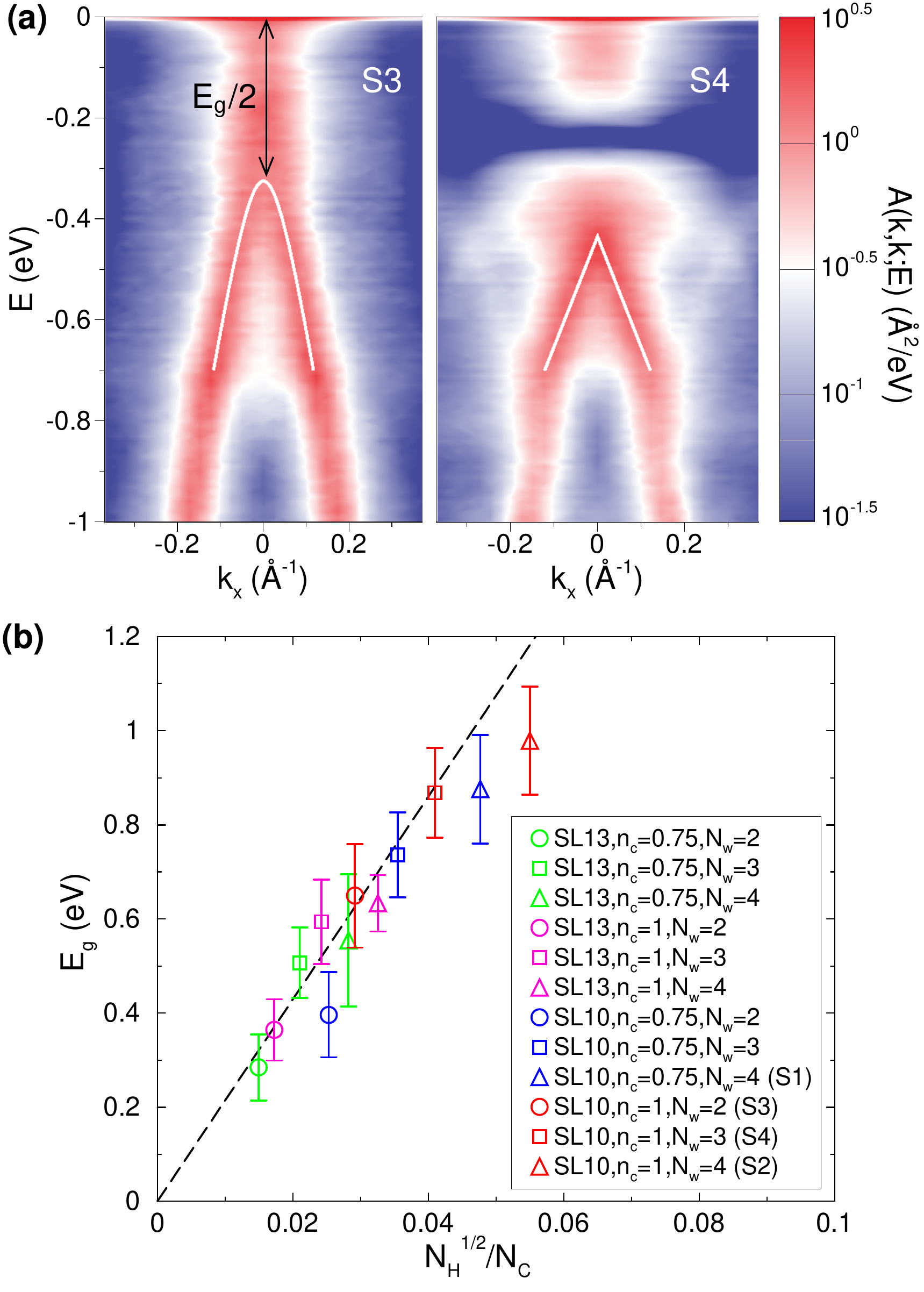}
\caption{\label{fig_bandgap}Band-gap extraction. (a) Momentum-energy resolved density of states for two sets of hydrogenated samples on iridium substrate: S3 is obtained with  the model parameters $N_w = 2$, $n_c = 1$, while S4 with $N_w = 3$, $n_c = 1$. Different fitting curves are used (white lines), given by Eqs.~31 and 29 of Supp.~info, for S3 and S4, respectively. The band gap is extracted with respect to the fitting curve. (b) Band gap extracted for the various sets of samples and plotted as a function of $\sqrt{N_H}/N_C$, where $N_H$ and $N_C$ are the average number of hydrogen and carbon atoms in the half-supercell, respectively. The fitting functions used for the extraction are listed, for each set, in Table~1 of Supp.~info. SL10 stands for graphene on iridium substrate (supercell made of 10$\times$10 graphene unit cells, see Fig.~\ref{fig_structure_arpes}a), while SL13 refers to graphene on a fictitious substrate (supercell made of 13$\times$13 graphene unit cells). The error bar is a measure of the broadening of the $A(\mathbf{k},\mathbf{k};E)$ plot. The dashed line is the curve $E_g=2 \hbar v_F \pi/\Delta$, with $v_F = (3/2)a_{\mathrm{CC}}|\gamma|/\hbar$ the Fermi velocity in pristine graphene and $\Delta = \sqrt{A_s/2} \left(N_C/\sqrt{N_H}\right)$, where $a_{\mathrm{CC}}$ is the carbon-carbon distance and $A_s = a_{\mathrm{CC}}^2 3\sqrt{3}/2$ the area of the unit cell of pristine graphene.}
\end{figure}

\textbf{Scaling of energy gap:} The energy gap is extracted from the momentum-energy resolved density of states for different sets of samples, corresponding to different values of cluster size, filling factor, and supercell size. The supercell size is increased by considering a fictitious substrate other than iridium: for fixed cluster size, this corresponds to increasing the neck width. Fig.~\ref{fig_bandgap}a illustrates the fitted band edges from $A(\mathbf{k},\mathbf{k};E)$, with details in Supp. info. An apparent universal scaling relation for the band gap is obtained when we plot the extracted band-gap values $E_g$ (together with a measure of the broadening of each $A(\mathbf{k},\mathbf{k};E)$ plot as error bar) against the quantity $\sqrt{N_H}/N_C$, where $N_H$ and $N_C$ are the average number of hydrogen and carbon atoms in the half-supercell (Fig.~\ref{fig_bandgap}b) \footnote{For constant supercell size and filling factor, the points seem to depart from the linear trend. This could be explained by the fact that, for some ensembles, the amount of disorder on the cluster edges is not strong enough to completely wash out the interference effect of the specific cluster shape. This situation is particularly evident for $N_w=3$, where the repeated band structure of the superlattice is still distinguishable in the $A(\mathbf{k},\mathbf{k};E)$ plot (see Fig.~\ref{fig_bandgap}a-right and additional plots in Supp. info).}. A similar relation also applies for the case of triangular graphene nanomesh \cite{pedersen08}. In Ref.~\onlinecite{petersen11}, it was stated that a universal relation does not hold for honeycomb graphene nanomesh. However, Fig.~\ref{fig_bandgap}b suggests that when disorder is included in the simulations, the scaling law $E_g = c \sqrt{N_H}/N_C$, with $c$ a constant, can be valid at low defect coverage for honeycomb superlattices as well. This is similar to the case of graphene nanoribbons, where theoretically the band gap depends on the precise atomic configuration \cite{son06}, while a general law $E_g \propto 1/W$, with $W$ the ribbon width, is always observed in the experiments \cite{han07} and commonly attributed to disorder \cite{gunlycke07,han10}. Regarding the proportionality constant, we found that the expression $c = 2 \hbar v_F \pi / \sqrt{A_s/2}$, with $v_F$ the graphene Fermi velocity and $A_s$ the area of the graphene unit cell, fits fairly well the numerical data (dashed line in Fig.~\ref{fig_bandgap}b). We note that, by defining $\Delta = \sqrt{A_s/2} \left(N_C/\sqrt{N_H}\right)$, the scaling relation takes the form $E_g=2 \hbar v_F \pi/\Delta$. This equation can be thought of as arising from the quantization of the graphene dispersion relation $E = \pm \hbar v_F |\mathbf{k}|$ ($\mathbf{k}$ is here the wave vector around the $K$ point) with $|\mathbf{k}| = \pi/\Delta$ (1D quantization in random directions): $\Delta$ can thus be interpreted as an effective confinement length.\\

\begin{figure*}[t]
\includegraphics[scale=0.492]{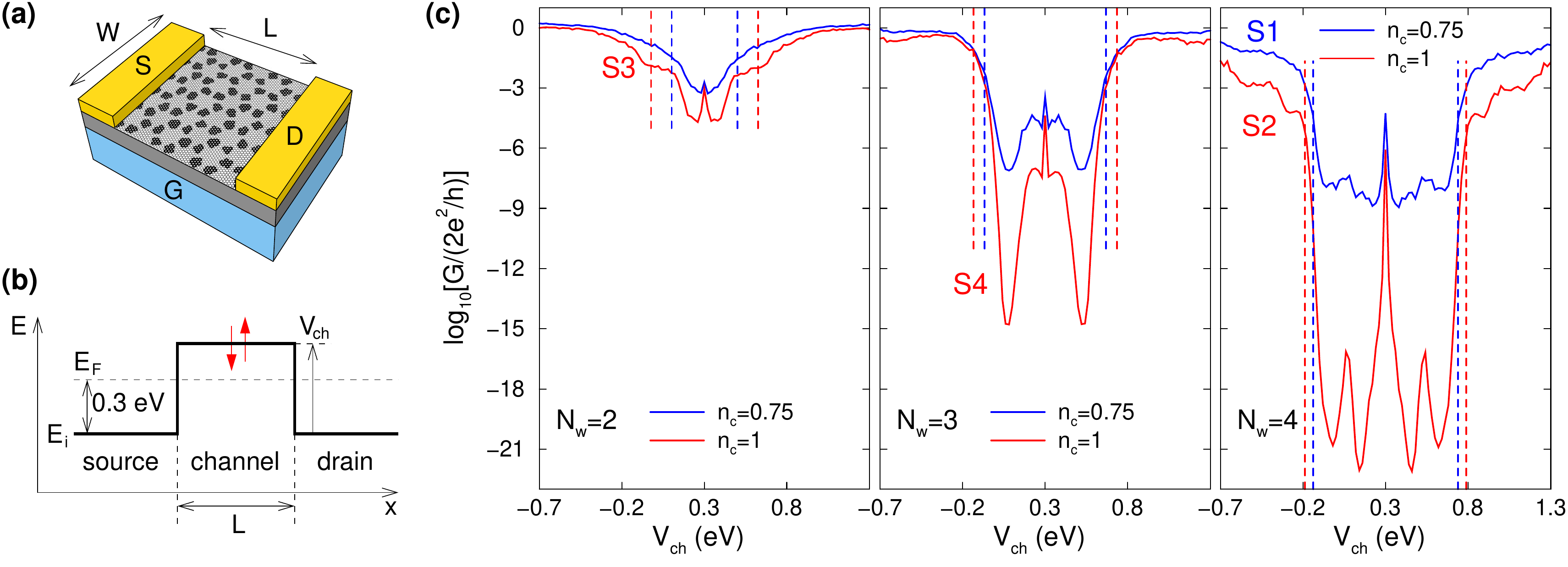}
\caption{\label{fig_conductance}Transport simulations. (a) Conceptual device under investigation: pattern-hydrogenated graphene is transferred to an insulating substrate and used as channel material of a field-effect transistor. (b) Profile of the potential energy used to simulate the structure in (a): the Fermi level in the leads $E_F$ is kept fixed, while the barrier height $V_{\mathrm{ch}}$ is varied to reproduce the effect of the back gate. Pristine graphene is used for the leads. (c) Zero-temperature conductance vs. $V_{\mathrm{ch}}$ for various sets of samples with $W=L=30$~nm and iridium substrate (SL10). From left to right, the cluster size, i.e. $N_w$, is increased; within the same plot, the cluster concentration $n_c$ is varied. 100 samples are considered for each set and the average is done on the logarithm of the normalized conductance (a motivation for this type of averaging can be found in \cite{anderson80}). The vertical lines indicate the band gap from Fig.~\ref{fig_bandgap}b.}
\end{figure*}
\begin{figure*}[t]
\includegraphics[scale=0.492]{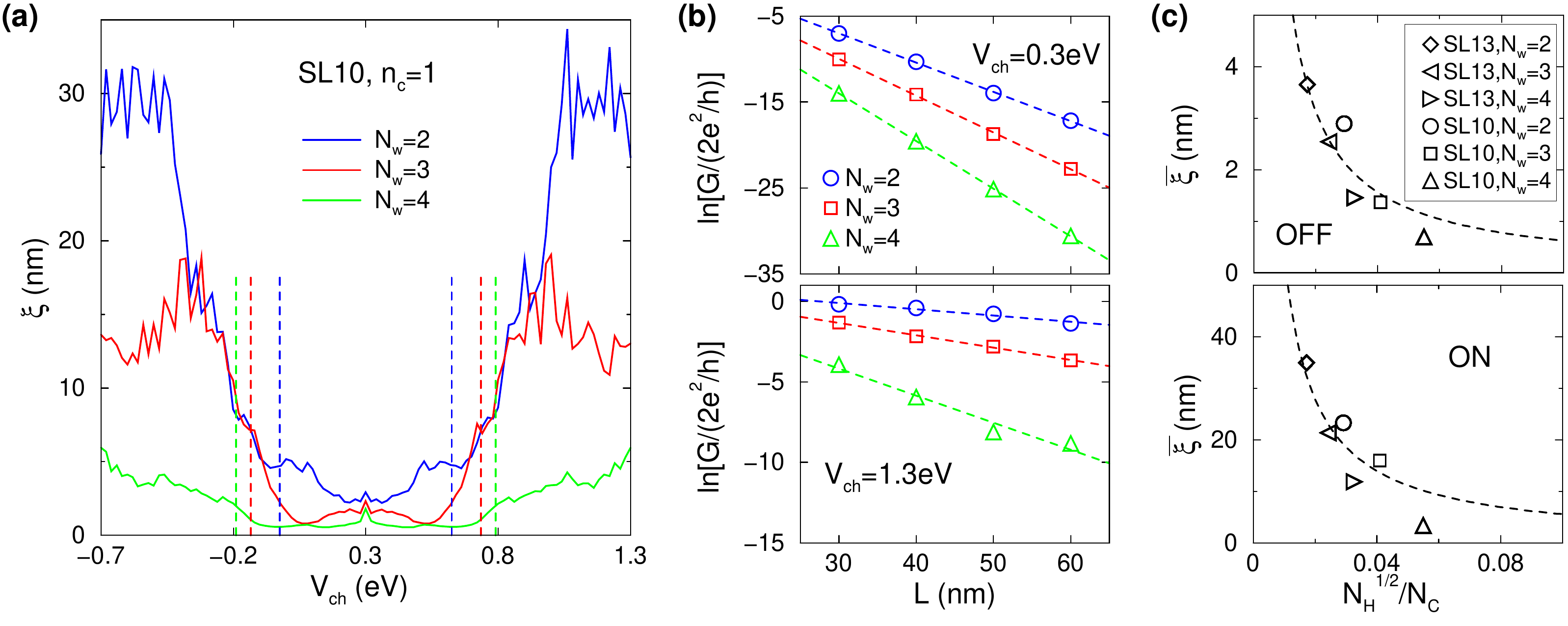}
\caption{\label{fig_localization}Decay length extraction. (a) Decay length vs. $V_{\mathrm{ch}}$ for sets of samples with different cluster size and fixed supercell size (SL10) and cluster concentration ($n_c=1$). The vertical lines indicate the band gap from Fig.~\ref{fig_bandgap}b. (b) Example of the decay length extraction at two different $V_\mathrm{ch}$ points. The dashed lines indicate the fitting with the formula $\ln [G/(2e^2/h)] = \ln g_0 - L / \xi$. (c) Average value of the decay length in the ``off'' and ``on'' state for various sets of samples with different supercell and cluster sizes, plotted as a function of $\sqrt{N_H}/N_C$. The dashed lines indicate the fitting curve $\bar{\xi} \propto N_C/\sqrt{N_H}$. The off state is defined as the bias region $|V_{\mathrm{ch}}-E_F| < E_G/2-B$, where $B$ is the half-broadening from Fig.~\ref{fig_bandgap}b, while the on state as 0.65 eV $<|V_{\mathrm{ch}}-E_F| <$ 0.75 eV.}
\end{figure*}

\textbf{Transport gap:} Next, we examine the electronic transport properties of pattern-hydrogenated graphene. Techniques for the transfer of graphene grown on metal surfaces to an insulating substrate have recently been developed \cite{li2_09}. We consider a three-terminal structure as shown in Fig.~\ref{fig_conductance}a and aim at predicting its low-temperature, low-bias conductance. Fig.~\ref{fig_conductance}b illustrates the potential energy along the device. The potential energy in the source and drain leads, as a result of metal induced doping, is kept fixed with respect to the Fermi level $E_F$. The channel potential $V_\mathrm{ch}$ is modulated by the back gate. Graphene is aligned with its armchair direction along the longitudinal direction of the device, in order to avoid edge transport effects. Only the channel is hydrogenated while the leads are pristine graphene. The conductance is computed by using the standard Green's function technique \cite{datta97} combined with a modified version of the algorithm described in Ref.~\onlinecite{sancho85}, which is commonly used for the calculation of the lead self-energies (see Supp. info). 

Fig.~\ref{fig_conductance}c shows the simulated, ensemble averaged zero-temperature conductance $G$ vs. $V_\mathrm{ch}$. The device size is kept fixed at $W=L=30$ nm, while different sets of hydrogenated samples are considered. It can be seen that patterned hydrogenation leads to a clear transport gap, increasing with $N_w$ and $n_c$. Also, the transport simulations agree well with our band-structure results: the transport gap matches the band gap from the $A(\mathbf{k},\mathbf{k};E)$ fitting (as indicated by vertical lines in Fig.~\ref{fig_conductance}c) and the peaks in the transport gap region correspond to the gap states in $A(\mathbf{k},\mathbf{k};E)$ \footnote{We note that, since the transport simulations confirm the previously extracted gap size, the gap is expected to have the same extension in other directions than the one used for Figs.~\ref{fig_structure_arpes}c, \ref{fig_bandgap}a.}. The $G$ vs. $V_\mathrm{ch}$ curve appears symmetrical, unlike the case for pristine graphene \cite{tony09}. This suggests that scattering is dominated by the channel, instead of the tunneling resistance due to $pn$ junctions.  \\



\textbf{Scaling of transport coefficients:} Finally, we examine how $G$ scales with $L$. Here, we consider devices with filling factor $n_c = 1$. The conductance is found to scale as $G \propto \exp(-L/\xi)$, where $\xi$ is a decay length. Fig.~\ref{fig_localization}a plots the extracted $\xi$ as function of $V_\mathrm{ch}$ bias, while Fig.~\ref{fig_localization}b illustrates the extraction of $\xi$ for two particular $V_\mathrm{ch}$ values. Next, we extract the average value $\bar{\xi}$ of the decay length in the ``off'' and ``on'' state and plot it against $\sqrt{N_H}/N_C$ as we have done previously for the band gap (Fig.~\ref{fig_localization}c, see caption for the definition of the ``on'' and ``off'' states).  One observes that, for almost all the samples, the value of the decay length in the ``off'' state is about an order of magnitude smaller than the corresponding value in the ``on'' state. For both cases, the average decay length seems to follow the general scaling law $\bar{\xi} \propto N_C/\sqrt{N_H}$ at low to moderate hydrogenation concentrations, albeit the ``off'' state exhibits a smaller proportionality constant. We note that in the ``off'' state the exponential decrease of $G$ with $L$ can be explained as evanescent transport through a clean band gap \cite{gunlycke07}; the scaling law $\bar{\xi} \propto \Delta \propto 1/E_g$ is in agreement with this interpretation. In the ``on'' state instead, the exponential decrease of $G$ with $L$ is an effect of quantum localization due to disorder and $\xi$ takes the meaning of a localization length \cite{lee85}. The scaling law $\bar{\xi} \propto \Delta$ is here less clear and an exponential dependence could also be possible, as suggested by recent experiments \cite{jaiswal11}. Moreover, dephasing effects, which are ignored in our simulations, could restore a diffusive transport regime, $G \propto 1/L$, as the temperature is raised.\\


\textbf{Conclusions:} In conclusion, a simple model for pattern-hydrogenated graphene was presented. Similar features are observed in the calculated $k$-resolved density of states in energy and in the experimental ARPES. The scaling of the energy gap on the parameters $N_C$ and $N_H$ was presented, including its electronic transport properties at low temperature. Our results indicate that pattern-hydrogenated graphene is a promising approach to the engineering of graphene nanomeshes with extremely scaled cluster sizes and neck widths.


\textbf{Acknowledgement:} The authors gratefully acknowledge support from Network for Computational Nanotechnology for computational services.


\newpage
\appendix

\section{Triangular versus honeycomb superlattice}

\begin{figure}[b]
\begin{center}
\includegraphics[scale=0.5]{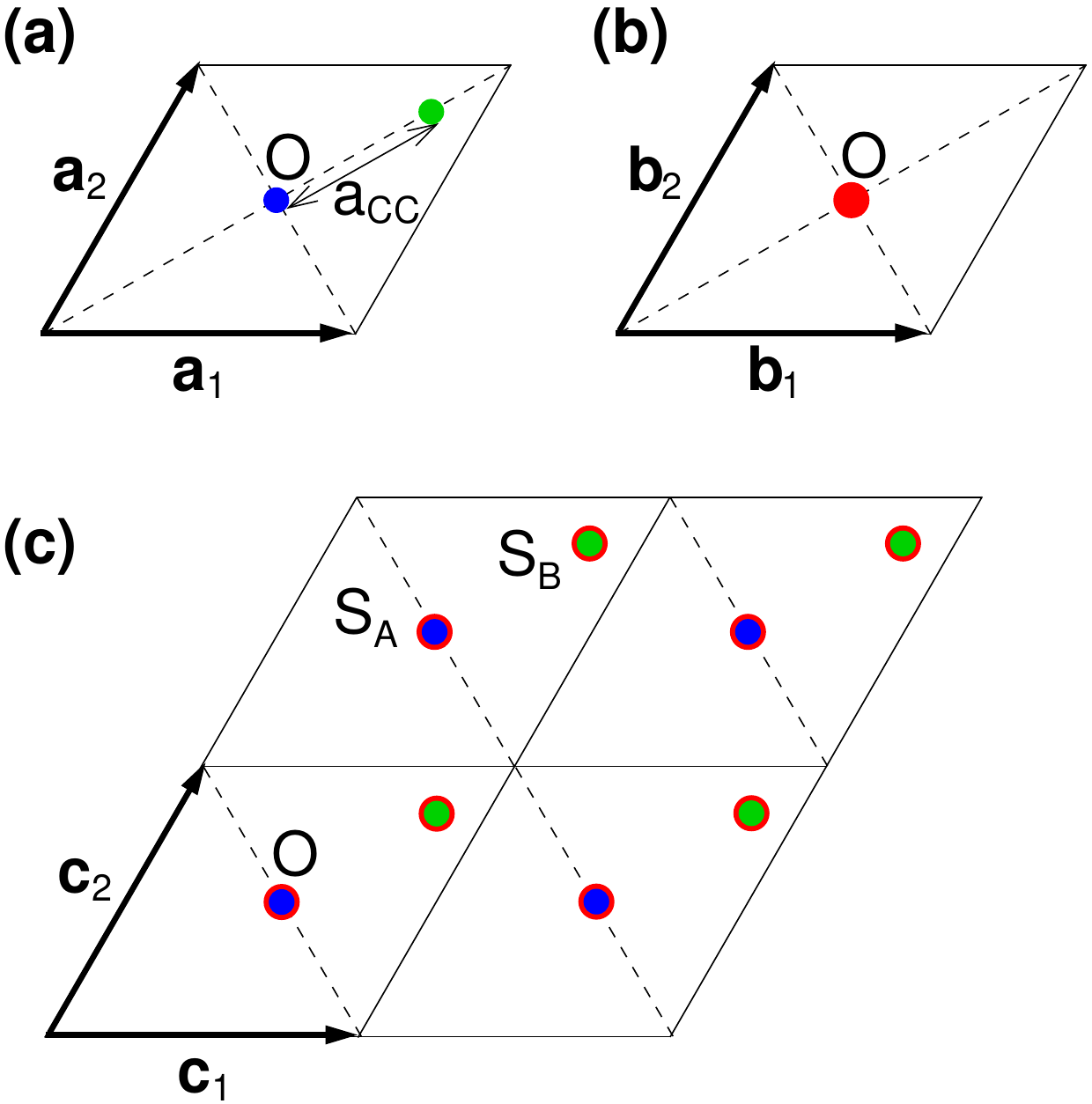}
\caption{\label{fig_latticesketch}(a) Graphene unit cell: the two carbon atoms are indicated with different colors. (b) Substrate unit cell. (c) Honeycomb superlattice generated by the superposition of the graphene and substrate lattices. $S_A$ ($S_B$) is the point inside the supercell where a carbon atom of the $A$ ($B$) graphene sublattice sits on top of a substrate atom.}
\end{center}
\end{figure}

While patterned hydrogenation has been experimentally demonstrated only for graphene on iridium, other substrate materials that can accomodate a monolayer graphene on their surface, e.g. rutenium \cite{deparga08}, could in principle be used. The resulting superlattice can be of the triangular or honeycomb type. Let $m$ and $n$ be the size of the supercell in units of the graphene and substrate lattice constants, respectively. Here it is demonstrated that a honeycomb superlattice is obtained whenever $m/n = (3M+1)/(3N)$ with $M, N \in \mathbb{Z}^+$. 

Figs.~\ref{fig_latticesketch}a and \ref{fig_latticesketch}b show the unit cell for the honeycomb graphene lattice and the typical triangular substrate surface layer, respectively. According to the definition of $m$ and $n$ given above, we assume that an $m \times m $ graphene supercell is commensurate with an $n \times n$ substrate supercell. Therefore, denoting the lattice vectors of graphene, substrate, and superlattice by $\mathbf{a}_{j}$, $\mathbf{b}_{j}$, and $\mathbf{c}_{j}$ ($j = 1, 2$), respectively, we have
\begin{equation}
\mathbf{c}_{j} = m \mathbf{a}_{j} = n \mathbf{b}_{j} \, .
\label{a1}
\end{equation}
Each pair $(m,n)$ and its multiples map to a unique composite system, e.g. $(4,3)$ is the same as $(8,6)$. Hence, we consider only the case where $m$ and $n$ are prime to each other, so that $\mathbf{c}_{j}$ are the primitive lattice vectors of the superlattice. We further assume that, at some point $O$ inside the supercell, a carbon atom sits directly on top of a substrate atom: such arrangement was found to be an energetically stable configuration \cite{ndiaye06}. We indicate the superlattice points that are equivalent to $O$ as $S_A$.

As shown by Fig.~\ref{fig_latticesketch}c, in order to generate a honeycomb superlattice, there must be another point $S_B$ inside the supercell where a carbon atom belonging to the opposite sublattice sits on top of a substrate atom. Also, for the superlattice to be regular, it can be derived that the distance between $S_{A}$ and $S_{B}$ must be equal to $\left|\mathbf{c}_{1}+\mathbf{c}_{2}\right|/3$. Inspecting the graphene lattice tells us that $S_{A}$ and $S_{B}$ have to be separated by a distance of $(3M+1)a_{\textrm{CC}}$, where $M\in\mathbb{Z}^{+}$ and $a_{\textrm{CC}}$ is the carbon-carbon distance. Hence, we can write
\begin{eqnarray}
\frac{1}{3}\left|\mathbf{c}_{1}+\mathbf{c}_{2}\right|=\frac{m}{3}\left|\mathbf{a}_{1}+\mathbf{a}_{2}\right|=(3M+1)a_{\textrm{CC}} \quad \Rightarrow \quad m=3M+1,  \quad   M\in\mathbb{Z}^{+} \, .
\label{a2}
\end{eqnarray}
In a similar fashion for the substrate, the $S_{A}$ and $S_{B}$ have to be separated by a distance of $Na_{\textrm{SS}}$, where $N\in\mathbb{Z}^{+}$ and $a_{\textrm{SS}}$ is the interatomic distance of the substrate. We can then write
\begin{eqnarray}
\frac{1}{3}\left|\mathbf{c}_{1}+\mathbf{c}_{2}\right|=\frac{n}{3}\left|\mathbf{b}_{1}+\mathbf{b}_{2}\right|=Na_{\textrm{SS}} \quad \Rightarrow \quad n=3N,  \quad   N\in\mathbb{Z}^{+} \, .
\label{a3}
\end{eqnarray}

In conclusion, the superlattice with the similar honeycomb structure as graphene, shown in Fig.~\ref{fig_latticesketch}c, can be generated by satisfying Eqs.~\ref{a2}-\ref{a3}. Otherwise, the superlattice would produce a triangular structure instead, with only repeated units of $S_{A}$. Examples of honeycomb superlattices are the cases $m/n = 10/9$ (graphene on iridium, also indicated as SL10 in the manuscript) and $m/n = 13/12$ (indicated as SL13 in the manuscript). 

\section{Model for patterned hydrogenation}

The model for generating the hydrogen clusters is described below.

Given a sample of pristine graphene on a certain substrate, we divide the structure in supercells, the supercell subdivision being chosen so that each supercell contains one $S_A$ and one $S_B$ point in symmetric positions (see Fig.~1a of the main text for an illustration: the centers of the two circles correspond to an $S_A$ or $S_B$ point). Each carbon atom can be denoted by the pair of indexes $(l,i)$, where $l$ is the index of the half-supercell to which it belongs and $i$ is the atom index inside the whole supercell. We introduce a binary random variable $Z_{l,i} \in \{0,1\}$ for each carbon atom: the atom is hydrogenated when $Z_{l,i} = 1$. $Z_{l,i}$ in turn is written as the product of other two binary random variables $X_l$ and $Y_i$, whose probability distributions are given below.

$Y_i$, controls the cluster formation inside each supercell. We propose the following formula for the probability $P(Y_i = 1)$:
\begin{equation} \label{eq_prob_p}
P(Y_i = 1) \equiv f(w_i) \, , \quad f(w_i = 0) = 0\, , \quad \frac{d f}{d w_i} \ge 0\, ,
\end{equation}
$w_i$ being a quantity defined for each carbon atom as
\begin{equation} \label{eq_prob_w}
w_i = \frac{d_i \left| d_i - \frac{1}{3} \sum_{\langle j \rangle} d_j - c\right|}{a_{\textrm{CC}}^2} \, ,
\end{equation}
where $d_i$ is the $xy$-plane distance between the carbon atom of index $i$ and its nearest-neighbor substrate atom (let $\mathbf{r}$ be the position vector in the $xy$-plane parallel to the surface),
\begin{equation}
d_i = \min_k \left| \mathbf{r}_{C_i} - \mathbf{r}_{S_k} \right| \, ,
\end{equation}
the summation over $j$ is restricted to the three carbon atoms that are nearest neighbor to the carbon atom of index $i$, and $c$ is simply a constant,
\begin{equation}
c = \frac{a_{\textrm{SS}}}{\sqrt{3}}-\sqrt{\frac{a_{\textrm{SS}}^2}{3}+a_{\textrm{CC}}^2-\frac{a_{\textrm{SS}} a_{\textrm{CC}}}{\sqrt{3}}} \, .
\end{equation}

\begin{figure}[t]
\begin{center}
\includegraphics[scale=0.5]{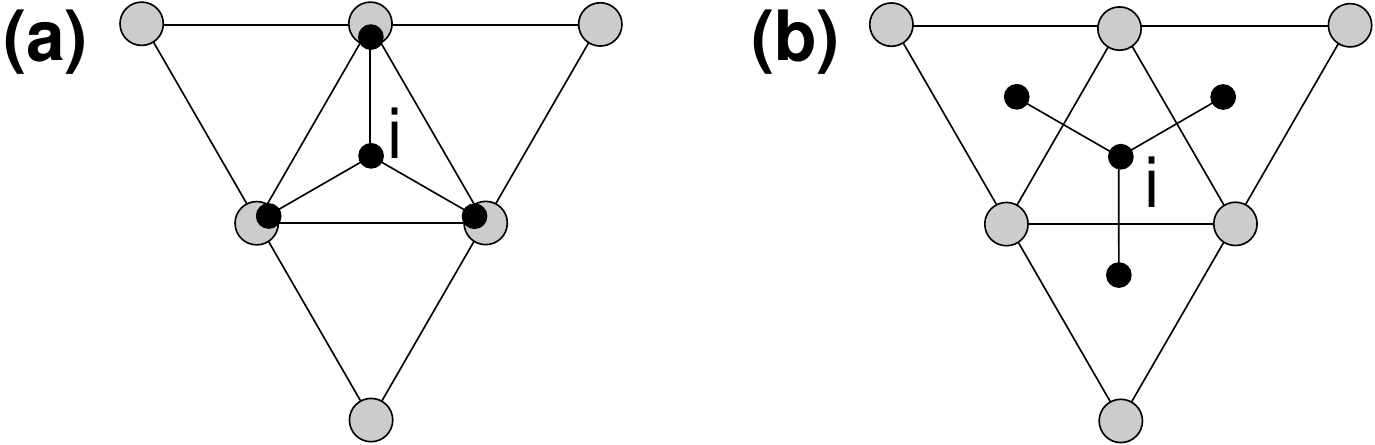}
\caption{\label{fig_best_worst}Best (a) and worst (b) cases for the probability of hydrogenation of a carbon atom of index $i$ located at a distance $d_i = a_{\textrm{SS}} / \sqrt{3}$  from the nearest-neighbor substrate atoms. Carbon (substrate) atoms are represented with black (gray) balls.}
\end{center}
\end{figure}

Eqs.~\ref{eq_prob_p}--\ref{eq_prob_w} can be justified by the following considerations. Experimentally, the hydrogen clusters tend to form around the regions where one graphene sublattice is located on top of substrate atoms, while the other sublattice is far from substrate atoms and can bind to hydrogen atoms on the opposite face \cite{balog10}. This translates in two conditions for the generic carbon atom to be hydrogenated. First, it should be located in between substrate lattice sites. The probability for adsorption whould then increase as its distance $d_i$ from the nearest-neighbor substrate atom increases. This effect is captured by the prefactor in (\ref{eq_prob_w}). However, if the considered carbon atom is located at the maximum distance from substrate atoms, equal to $a_{\textrm{SS}} / \sqrt{3}$, the probability for adsorption should distinguish between the case in which the three nearest-neighbor carbon atoms are located close to substrate atoms (high probability, Fig.~\ref{fig_best_worst}a) and the case in which also the three nearest neighbors are between substrate atoms (low probability, Fig.~\ref{fig_best_worst}b). We can note that in the first case $d_i \gg \frac{1}{3} \sum_{\langle j \rangle} d_j$, while in the second case $d_i \sim \frac{1}{3} \sum_{\langle j \rangle} d_j$. This explains the second factor in (\ref{eq_prob_w}), where the constant $c$ serves only to set the probability to 0 for the worst case (Fig.~\ref{fig_best_worst}b).

\begin{figure}[t]
\begin{center}
\includegraphics[scale=0.75]{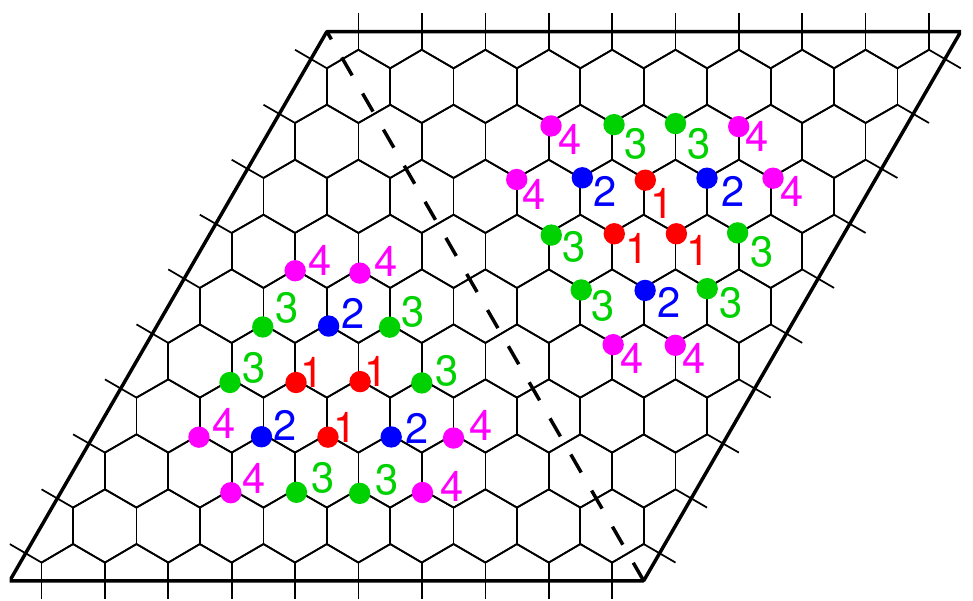}
\caption{\label{fig_supercell_w}Supercell of graphene on iridium: location of the carbon atoms with the four largest values of $w_i$, i.e. $w_i = w^{(1)}$, $w_i = w^{(2)}$, etc.}
\end{center}
\end{figure}

In (\ref{eq_prob_p}) we have omitted the actual functional dependence of $P(Y_i = 1)$ on $w_i$. Since this relationship depends on the physical hydrogenation process and it is unknown, we choose here a simple cut-off model. For a given superlattice unit cell, all the possible values of $w_i$ are computed and labeled in decreasing order as $w^{(1)}, w^{(2)}, \ldots$ (the location of the corresponding carbon atoms is shown in Fig.~\ref{fig_supercell_w} for the case of iridium substrate). Then, the probability $P(Y_i = 1)$ is assigned as
\begin{eqnarray}
P(Y_i = 1) \equiv f(w_i) = \left\{
\begin{array}{ll}
1   & \mbox{if $w_i = w^{(j)}$ with $j < N_w$,} \\
0.5 & \mbox{if $w_i = w^{(j)}$ with $j = N_w$,} \\
0   & \mbox{if $w_i = w^{(j)}$ with $j > N_w$.}
\end{array}
\right.
\end{eqnarray}
With this method, a cluster of hydrogen atoms is formed around the sites where the quantity $w_i$ tends to grow (i.e. around $S_A$ and $S_B$). The input parameter $N_w$ controls the size of this cluster. The disorder is only located at the cluster edges.

$X_l$, instead, is used to generate the superlattice disorder, which consists in some hydrogen clusters being randomly missing from the superlattice. The probability $P(X_l = 1)$ is set equal to the input parameter $n_c$, with $0 \le n_c \le 1$, which therefore assumes the meaning of the ratio between the average number of hydrogenated half-supercells (or equivalently average number of clusters) and the total number of half-supercells. 

The hydrogenation model described above produces clusters inside which only one graphene sublattice is hydrogenated. This leads to the formation of midgap states in the electronic structure, associated with dangling bonds. However, these states are believed to be an artefact of the TB model, due to the fact that bond relaxation is neglected. To avoid this, after hydrogen atoms are generated according to the method described above, a final step is introduced: additional hydrogen atoms are placed on top of the carbon atoms that have at least two nearest neighbors being hydrogenated.

\section{Calculation of momentum-energy resolved density of states}

\begin{figure}[htp]
\begin{center}
\includegraphics[scale=0.75]{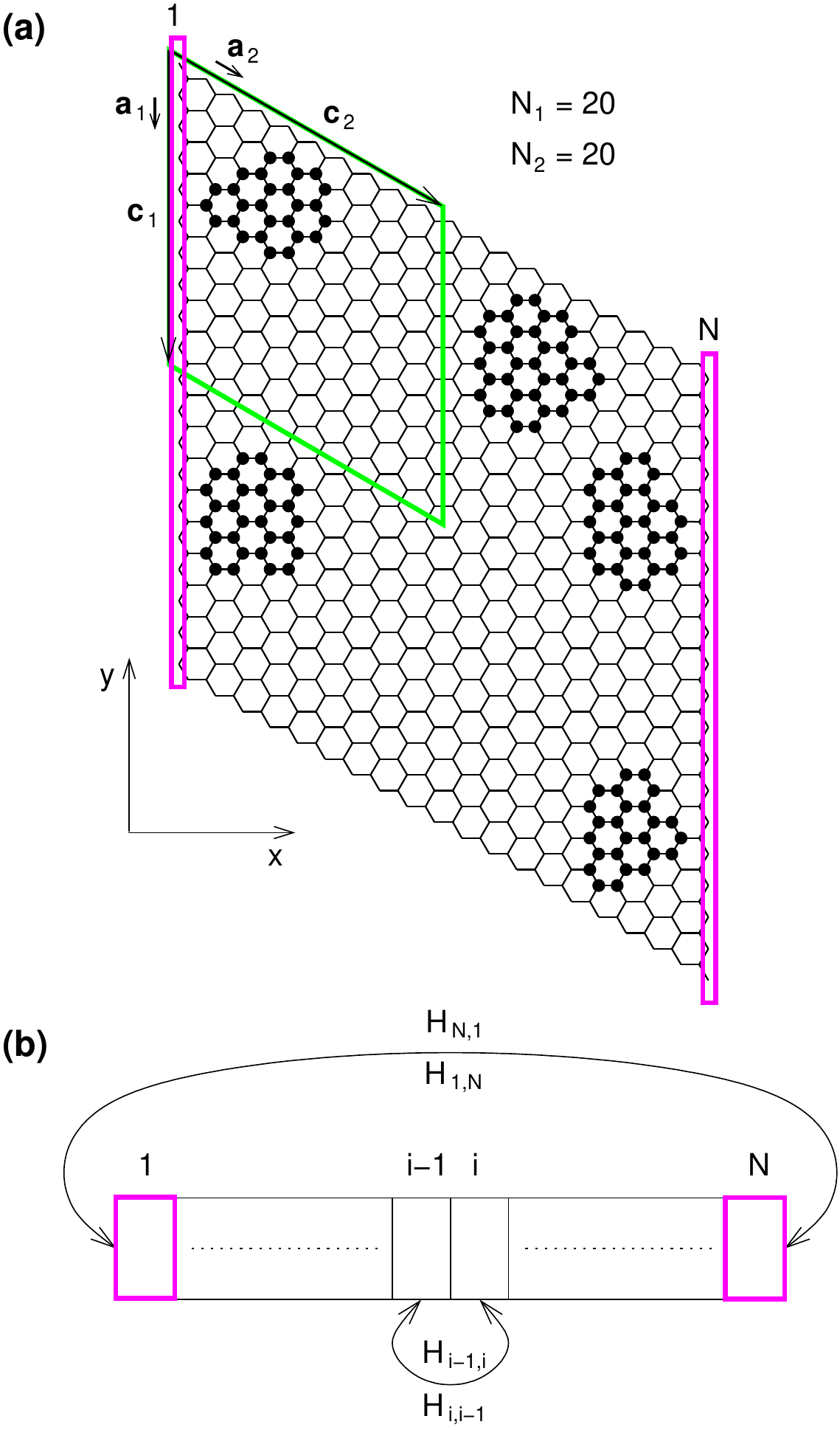}
\caption{(a) Example of sample used for calculating the momentum-energy resolved density of states. The structure is made of $N_1 \times N_2$ graphene unit cells, with $N_1$ and $N_2$ being chosen so that the sample is a multiple of the superlattice unit cell (region enclosed by the green line). Periodic boundary conditions are applied on both the $\mathbf{a}_1$ and $\mathbf{a}_2$ directions. (b) The same structure can be viewed as a linear chain of $N = 2 N_2$ slabs (each slab corresponding to a row of atoms) with a periodic closure at the two ends.} \label{fig_periodic_structure}
\end{center}
\end{figure}

We consider a sample composed of $N_1$ and $N_2$ graphene unit cells along the directions of $\mathbf{a}_1$ and $\mathbf{a}_2$, respectively, with periodic boundary conditions on both directions. $N_1$ and $N_2$ are chosen to be multiples of $m$, the size of the supercell in units of the graphene lattice constant, so that the sample contains exactly $\frac{N_1}{m} \times \frac{N_2}{m}$ supercells. The sample is then hydrogenated as described in the previous section. An example, obtained with $N_1 = N_2 = 20$, $m/n=10/9$ (graphene on iridium), $n_c = 0.75$, and $N_w=4$, is shown in Fig.~\ref{fig_periodic_structure}a. The actual samples that are simulated are much larger: $N_1 = N_2 = 120$ is used for the SL10 case, while $N_1 = N_2 = 117$ for the SL13 case. Such values have been checked to be large enough to ensure the convergence of $A(\mathbf{k},\mathbf{k};E)$ (see next section).

The generic orbital of the TB representation can be indicated as $|\mathbf{l},q\rangle$, where $\mathbf{l}$ is the graphene lattice vector, i.e. $\mathbf{l}=\sum_i n_i a_i$ with $n_i = 1, \ldots N_i$, and $q$ is the orbital index within the cell ($q = 1, 2$ for the $A$ and $B$ carbon orbitals, respectively). This is the real space representation. However, one could also work in the $k$-space representation, which is obtained by restricting $q=1,2$ (projected space for carbon atoms only) and by using as basis the set $\{| \mathbf{k}, q \rangle\}$ defined by $\langle \mathbf{l}, q_1 | \mathbf{k}, q_2 \rangle = \delta_{q_1,q_2} e^{\mathrm{i} \mathbf{k} \cdot \mathbf{l}} / \sqrt{N_1 N_2}$. The $\mathbf{k}$ vectors are discrete because a finite volume is considered: if $\mathbf{d}_i$ are the primitive vectors of the reciprocal lattice, i.e. $\mathbf{a}_i \cdot \mathbf{d}_j = 2 \pi \delta_{i,j}$, we have $\mathbf{k} = (m_1/N_1) \mathbf{d}_1  + (m_2/N_2) \mathbf{d}_2$, where $m_1, m_2 \in \mathbb{Z}$ (it can be proved that the $K$ point is included in the grid if both $N_1$ and $N_2$ are multiple of 3). Also, since the set of $\mathbf{l}$ vectors is discrete, it follows that only a number $N_1 N_2$ of  $\mathbf{k}$ vectors, spanning a Brillouin zone, give rise to independent basis vectors.

Let $H$ be the electron Hamiltonian. We recall that the spectral function at the energy $E$ is defined as $A = \mathrm{i} (G^r-G^a)$, where $G^r = [(E+\mathrm{i}\eta)I-H]^{-1}$ is the retarded Green's function and $G^a = {G^r}^\dagger$ is the advanced one ($\eta$ is an infinitesimal positive quantity). While the diagonal elements of $A$ in real space have the meaning of a density of states in energy and physical space, its diagonal elements in $k$-space give the density of states in energy and momentum, which in turn corresponds to the physical quantity measured by ARPES. The diagonal elements of $A$ in $k$-space can be expanded as
\begin{align}\label{eq_akk_1}
A(\mathbf{k},q;\mathbf{k},q;E) &= \sum_{\mathbf{l}_1,q_1} \sum_{\mathbf{l}_2,q_2}
\langle \mathbf{k},q | \mathbf{l}_1,q_1 \rangle A(\mathbf{l}_1,q_1;\mathbf{l}_2,q_2;E) \langle \mathbf{l}_2,q_2 | \mathbf{k},q \rangle \nonumber \\
&= \frac{1}{N_1 N_2} \sum_{\mathbf{l}_1} \sum_{\mathbf{l}}
e^{- \mathrm{i} \mathbf{k} \cdot \mathbf{l}} A(\mathbf{l}_1,q;\mathbf{l}_1-\mathbf{l},q;E) \, ,
\end{align}
where we recognize a discrete Fourier transform with respect to the relative variable $\mathbf{l} = \mathbf{l}_1-\mathbf{l}_2$. We define
\begin{equation}\label{eq_akk_2}
A(\mathbf{k},\mathbf{k};E)= \frac{A_s}{(2 \pi)^2} \sum_q A(\mathbf{k},q;\mathbf{k},q;E) \, ,
\end{equation}
where $A_s = a_{\mathrm{CC}}^2 3\sqrt{3}/2$ is the area of the graphene unit cell, so that $A(\mathbf{k},\mathbf{k};E)/(2 \pi)$ gives the number of states per unit energy, per unit $\mathbf{k}$, and per graphene unit cell (apart from spin degeneracy). The calculation of $A(\mathbf{k},\mathbf{k};E)$ is thus performed by first computing the spectral function in real space and then Fourier transforming according to (\ref{eq_akk_1}-\ref{eq_akk_2}).

A method to compute $A$ in real space is by diagonalization of $H$. Indeed, if $\{| \psi_\alpha \rangle\}$ are the orthonormal eigenstates of $H$ with corresponding eigenvalues $\{\epsilon_\alpha\}$, we have
\begin{multline} \label{eq_arr}
A(\mathbf{l}_1,q_1;\mathbf{l}_2,q_2;E) = \sum_\alpha \frac{2 \eta}{(E - \epsilon_\alpha)^2 + \eta^2} \psi_\alpha(\mathbf{l}_1,q_1) \psi_\alpha^*(\mathbf{l}_2,q_2) 
\stackrel{\eta \rightarrow 0}{\longrightarrow}
2 \pi \sum_\alpha \delta(E -\epsilon_\alpha) \psi_\alpha(\mathbf{l}_1,q_1) \psi_\alpha^*(\mathbf{l}_2,q_2) \, ,
\end{multline}
with $\psi_\alpha(\mathbf{l},q) = \langle \mathbf{l}, q | \psi_\alpha \rangle$ the generic eigenfunction in real space. The numerical computation of (\ref{eq_arr}) can be efficiently done by setting a finite value of $\eta$ and by finding for each energy $E$ the eigenvalues (and corresponding eigenvectors) that are closest to it, using well known methods for large and sparse eigenvalue problems \footnote{\url{http://www.caam.rice.edu/software/ARPACK/}}. Nevertheless, we propose here an alternative method based on Green's functions. Although in the case considered here our method does not improve the computational time with respect to the diagonalization technique, because almost all the off-diagonal of the spectral function in real space need to be calculated, the method could be interesting in other situations, such as for the calculation of the local density of states in real space, where only few elements of the spectral function are needed. The method could also be useful when the storage of the eigenvectors is a major problem.

As illustrated in Fig.~\ref{fig_periodic_structure}a, we divide the sample in $N = 2 N_2$ slabs along the $\mathbf{a}_2$ direction so that each slab corresponds to a row of atoms (the choice between $\mathbf{a}_1$ and $\mathbf{a}_2$ is arbitrary). The structure is therefore of the type in Fig.~\ref{fig_periodic_structure}b: a linear chain of slabs with a periodic boundary conditions at the two ends. Using block matrix notation in real space, $\Omega = (E+\mathrm{i}\eta)I-H$ has the form
\begin{equation} \label{eq_eih_periodic}
\Omega = \left( \begin{array}{ccccc}
\Omega_{1,1} & \Omega_{1,2} &         &           & \Omega_{1,N}   \\
\Omega_{2,1} & \Omega_{2,2} & \Omega_{2,3} &           &           \\
        & \Omega_{3,2} & \ddots  & \ddots    &           \\
        &         & \ddots  & \ddots    & \Omega_{N-1,N} \\
\Omega_{N,1} &         &         & \Omega_{N,N-1} & \Omega_{N,N}
\end{array} \right) \, ,
\end{equation}
where each block represents the coupling between a pair of slabs. We notice that only the elements of the spectral function that connect orbitals belonging to the same graphene sublattice (i.e. same $q$) are needed in (\ref{eq_akk_1}-\ref{eq_akk_2}). From Fig.~\ref{fig_periodic_structure}a, it can be seen that there is a correspondence between $q=1,2$ and the odd and even slabs, respectively. Therefore, only the matrix block of $A$ (and thus of $G^r$) connecting slabs with the same parity need to be calculated. In order to avoid the calculation of the unnecessary matrix blocks, the renormalization method \cite{grosso89} is employed: an equivalent $\Omega^{\mathrm{odd}}$ ($\Omega^{\mathrm{even}}$) matrix for the odd (even) slabs alone is computed by decimating the even (odd) ones.
\begin{figure}[tp]
\begin{center}
\includegraphics[scale=0.85]{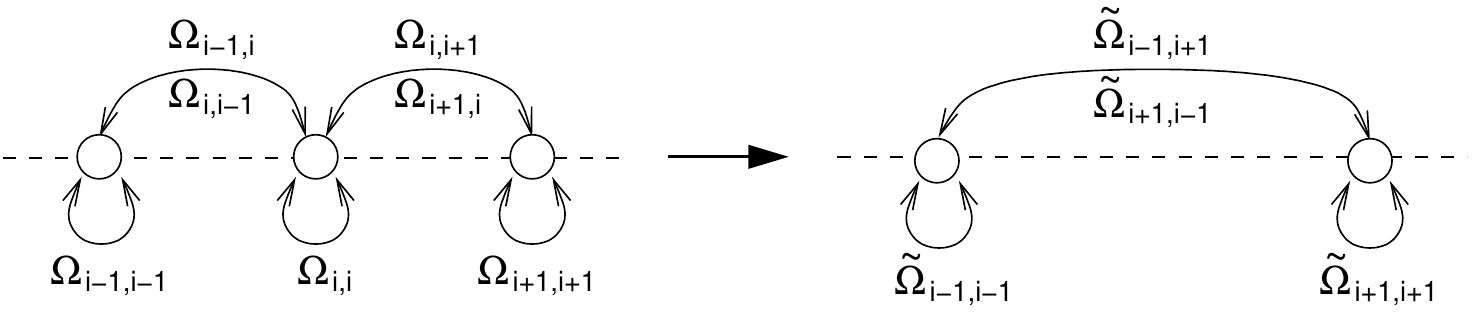}
\caption{Pictorial representation of the renormalization method in \cite{grosso89}: the circles represent nodes (or slabs) and the arrows the different types of coupling between them. The method consists in eliminating node $i$, while leaving unchanged the solution for all the remaining nodes by a proper renormalization of the matrix blocks of $\Omega$.} \label{fig_decimation}
\end{center}
\end{figure}
We recall that the decimation of a single node $i$ from a linear chain, as depicted in Fig.~\ref{fig_decimation}, is achieved by renormalizing the matrix blocks of $\Omega$ for the adjacent nodes according to the formulas \cite{grosso89}:
\begin{eqnarray} \label{eq_decimation}
\tilde{\Omega}_{i-1,i-1} \! &= &\!  \Omega_{i-1,i-1}-\Omega_{i-1,i} (\Omega_{i,i})^{-1} \Omega_{i,i-1} \, , \nonumber \\
\tilde{\Omega}_{i-1,i+1} \! &= &\! -\Omega_{i-1,i} (\Omega_{i,i})^{-1} \Omega_{i,i+1} \, , \nonumber \\
\tilde{\Omega}_{i+1,i-1} \! &= &\! -\Omega_{i+1,i} (\Omega_{i,i})^{-1} \Omega_{i,i-1} \, , \nonumber \\
\tilde{\Omega}_{i+1,i+1} \! &= &\! \Omega_{i+1,i+1}-\Omega_{i+1,i} (\Omega_{i,i})^{-1} \Omega_{i,i+1} \, .
\end{eqnarray}
By repeated use of (\ref{eq_decimation}), the following algorithm can be derived to compute the renormalized matrix $\Omega^{\mathrm{odd}}$ for the odd slabs alone (Fig.~\ref{fig_decimation_odd}):
\begin{figure}[tp]
\begin{center}
\includegraphics[scale=0.75]{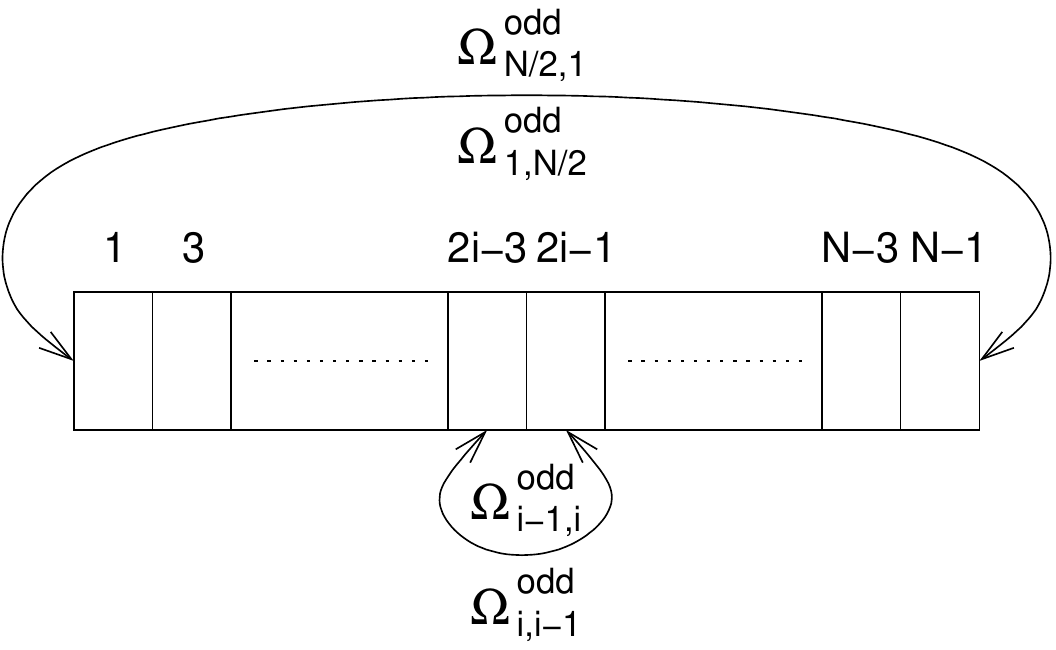}
\caption{Same structure as in Fig.~\ref{fig_periodic_structure}b, where the even slabs have been decimated.} \label{fig_decimation_odd}
\end{center}
\end{figure}
\begin{itemize}
\item[1.] for $i = 1, \ldots, N/2$ initialize
\begin{equation}
\Omega^{\mathrm{odd}}_{i,i} = \Omega_{2i-1,2i-1} \, ;
\end{equation}
\item[2.] for $i = 1, \ldots, N/2$
  \begin{itemize}
  \item set $j = 2i$, $k = \mathrm{mod}(i,\frac{N}{2})+1$, $l = \mathrm{mod}(j,N)+1$
  \item compute
  \begin{align}
  \Omega^{\mathrm{odd}}_{i,i} &= \Omega^{\mathrm{odd}}_{i,i}-\Omega_{j-1,j} (\Omega_{j,j})^{-1} \Omega_{j,j-1} \, , \nonumber \\
  \Omega^{\mathrm{odd}}_{k,k} &= \Omega^{\mathrm{odd}}_{k,k}-\Omega_{l,j} (\Omega_{j,j})^{-1} \Omega_{j,l} \, , \nonumber \\
  \Omega^{\mathrm{odd}}_{i,k} &= -\Omega_{j-1,j} (\Omega_{j,j})^{-1} \Omega_{j,l} \, , \nonumber \\
  \Omega^{\mathrm{odd}}_{k,i} &= -\Omega_{l,j} (\Omega_{j,j})^{-1} \Omega_{j,j-1} \, .
  \end{align}
  \end{itemize}
\end{itemize}
A similar algorithm can be derived for $\Omega^{\mathrm{even}}$. Both $\Omega^{\mathrm{odd}}$ and $\Omega^{\mathrm{even}}$ have the same shape as $\Omega$ (Eq.~\ref{eq_eih_periodic}), but with $N \rightarrow N/2$. In the following, we will therefore refer to $\Omega$ for brevity, implicitly assuming that what we say must be applied separately to $\Omega^{\mathrm{odd}}$ and $\Omega^{\mathrm{even}}$.

The retarded Green's function at the energy $E$ is obtained by inverting $\Omega$. Here we present an algorithm for recursively calculating the blocks of $G^r$, extending the one in \cite{lake97} for the case of $\Omega_{1,N}, \Omega_{N,1} \ne 0$. The algorithm is based on Dyson's equations. We recall that if the Hamiltonian is expressed as $H = H_0+H_1$  ($H_0$ is called the unperturbed Hamiltonian and $H_1$ the perturbation one), the Dyson equations are given by $G^r= G^r_0+G^r_0 H_1 G^r = G^r_0+G^r H_1 G^r_0$, where $G^r_0$ is the retarded Green's function corresponding to $H_0$. While the formulas that are presented here directly exploit time reversal symmetry, i.e. the fact that $G^r = (G^r)^T$, their extension to the general case is straightforward. 

\begin{figure}[t]
\begin{center}
\includegraphics[scale=0.75]{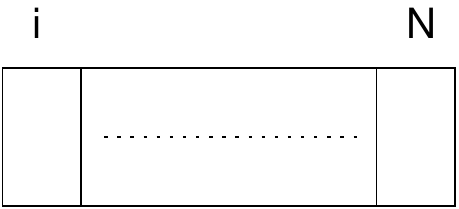}
\caption{Structure corresponding to $g^{r\mathcal{R},(i)}$.} \label{fig_structure_gr_right}
\end{center}
\end{figure}

The first part of the algorithm consists in the calculation of certain blocks of $g^{r\mathcal{R},(i)}$ for $i = 1, \ldots, N$, where $g^{r\mathcal{R},(i)}$ is the retarded Green's function corresponding to the structure composed of only the nodes from $i$ to $N$ without the periodic closure (Fig.~\ref{fig_structure_gr_right}).
Indeed, by applying Dyson's equations to the structure in Fig.~\ref{fig_structure_gr_right} with the coupling blocks $-\Omega_{i,i+1}$ and $-\Omega_{i+1,i}$ treated as the perturbation Hamiltonian, it is possible to relate $g^{r\mathcal{R},(i)}$ to $g^{r\mathcal{R},(i+1)}$ and derive the following equations:
\begin{itemize}
\item[1.] initialize
  \begin{equation}
  g^{r\mathcal{R},(N)}_{N,N} = (\Omega_{N,N})^{-1} \, ;
  \end{equation}
\item[2.] for $i = N-1, \ldots, 1$ compute
  \begin{align}
  g^{r\mathcal{R},(i)}_{i,i} &= (\Omega_{i,i} - \Omega_{i,i+1} g^{r\mathcal{R},(i+1)}_{i+1,i+1} \Omega_{i+1,i})^{-1} \, , \\
  g^{r\mathcal{R},(i)}_{i,N} &= - g^{r\mathcal{R},(i)}_{i,i} \Omega_{i,i+1} g^{r\mathcal{R},(i+1)}_{i+1,N} \, , \\
  g^{r\mathcal{R},(i)}_{N,i} &= (g^{r\mathcal{R},(i)}_{i,N})^T \, , \\
  g^{r\mathcal{R},(i)}_{N,N} &= g^{r\mathcal{R},(i+1)}_{N,N} - g^{r\mathcal{R},(i+1)}_{N,i+1} \Omega_{i+1,i} g^{r\mathcal{R},(i)}_{i,N} \, .
  \end{align}
\end{itemize}
The blocks $g^{r\mathcal{R},(i)}_{N,N}$ for $i > 1$ can be discarded. The second part of the algorithm is obtained by applying again Dyson's equations, but to the original structure in Fig.~\ref{fig_periodic_structure}b, with the perturbation Hamiltonian given by the coupling blocks $-\Omega_{i-1,i}$, $-\Omega_{i,i-1}$, $-\Omega_{N,1}$, and $-\Omega_{1,N}$. The formulas are as follows
\begin{itemize}
\item[3.] initialize
  \begin{multline}
  G^r_{1,1} = \left[ I + g^{r\mathcal{R},(1)}_{1,N} \Omega_{N,1} - g^{r\mathcal{R},(1)}_{1,1} \left( I + \Omega_{1,N} g^{r\mathcal{R},(1)}_{N,1} \right)^{-1} \Omega_{1,N} g^{r\mathcal{R},(1)}_{N,N} \Omega_{N,1} \right]^{-1} \times \\
  \times \left[ g^{r\mathcal{R},(1)}_{1,1} - g^{r\mathcal{R},(1)}_{1,1} \left( I + \Omega_{1,N} g^{r\mathcal{R},(1)}_{N,1} \right)^{-1} \Omega_{1,N} g^{r\mathcal{R},(1)}_{N,1} \right] \, ; \\
  \end{multline}
\item[4.] for $i = 1, \ldots, N$
  \begin{itemize}
  \item if $i>1$ compute
  \begin{equation}
  G^r_{i,i} = g^{r\mathcal{R},(i)}_{i,i} - g^{r\mathcal{R},(i)}_{i,i} \Omega_{i,i-1} G^r_{i-1,i} -
  g^{r\mathcal{R},(i)}_{i,N} \Omega_{N,1} G^r_{1,i} \, ,
  \end{equation}
  \item for $j = i+1, \ldots, N$ compute
  \begin{align}
  G^r_{i,j} &= - G^r_{i,j-1} \Omega_{j-1,j} g^{r\mathcal{R},(j)}_{j,j} -
  G^r_{i,1} \Omega_{1,N} g^{r\mathcal{R},(j)}_{N,j} \, , \label{eq_grij} \\
  G^r_{j,i} &= (G^r_{i,j})^T \label{eq_grji} \, .
  \end{align}  
  \end{itemize}
\end{itemize}
As soon as each block of $G^r$ is computed, its contribution to (\ref{eq_akk_1}) can be evaluated and the block is ready for being discarded unless is used later by the algorithm. It can be checked that the first row of $G^r$ has to be fully saved, while $G^r_{i,j}$ is needed to calculate $G^r_{i,j+1}$ and $G^r_{i,i+1}$ to calculate $G^r_{i+1,i+1}$. In conclusion, compared to the direct inversion of the matrix in (\ref{eq_eih_periodic}), the proposed algorithm allows to reduce of about a factor of 4 the number of blocks that are calculated, due to both the decimation of nodes with different parity and the exploitation of time reversal symmetry; moreover, the blocks are recursively calculated one after the other, thus saving memory. We note that our algorithm could also be used for partial inversion of (\ref{eq_eih_periodic}), as in the case of the calculation of the density of states in energy, where only the diagonal elements of the spectral function in real space are required: in this case, for $i>1$, (\ref{eq_grij}) can be limited to $j=i+1$ and (\ref{eq_grji}) is not required. Regarding the stability of the overall algorithm, it should be pointed out that a value of $\eta = 10^{-3}$~eV was necessary in the simulations to avoid numerical artefacts; however, the corresponding broadening introduced in the $A(\mathbf{k},\mathbf{k};E)$ plot is way much smaller than the one due to disorder.

\section{Convergence study w.r.t. sample size (and ensemble size)}

In order to account for disorder across different cells of the superlattice, each sample has to be large enough so that it contains a sufficient number of clusters and the effect of the periodic boundary condition is washed out. In addition, the size of the sample determines the discretization step in $k$-space (the total number of $\mathbf{k}$ points along the red line in the inset of Fig.~1c of the main text is equal to $N_2$): for a high-quality plot of $A(\mathbf{k},\mathbf{k};E)$, each sample has to be large enough so that the grid in $k$-space is sufficiently fine.

\begin{figure}[tp]
\begin{center}
\includegraphics[scale=0.72]{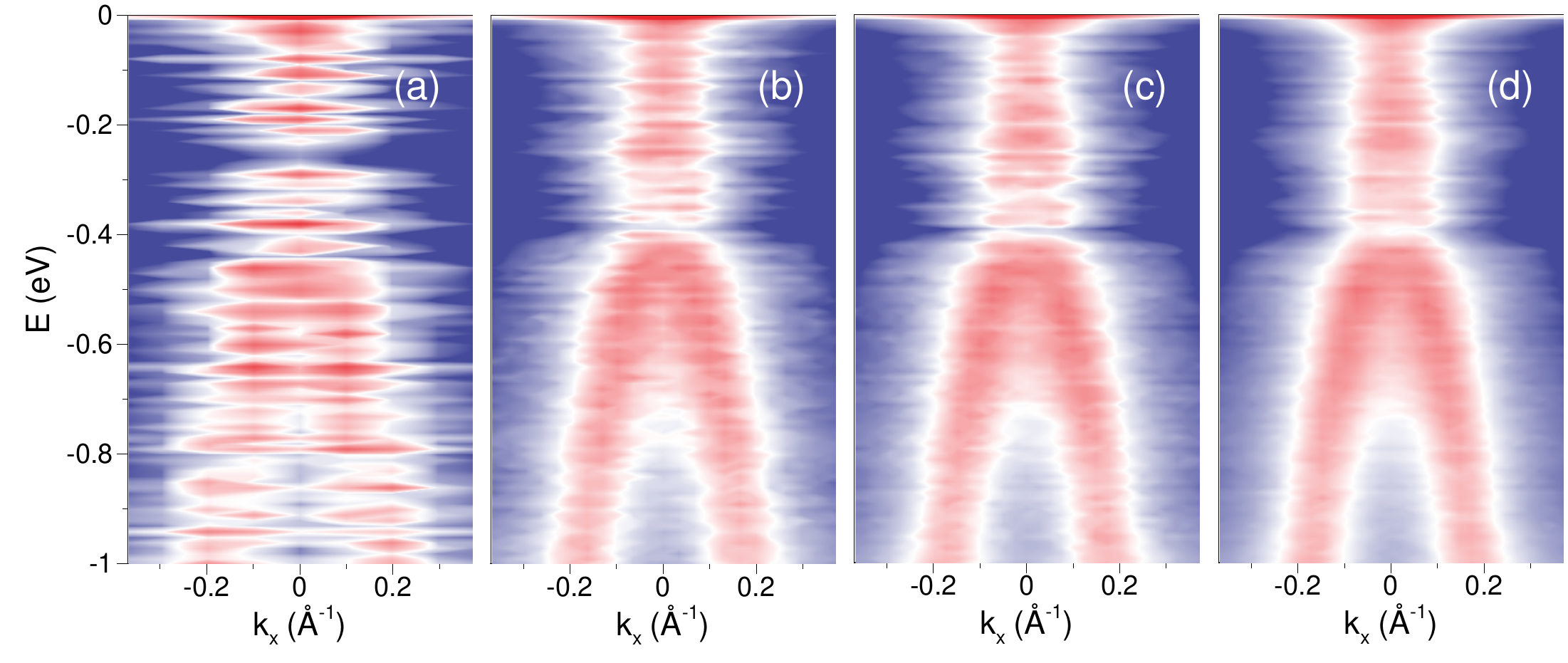}
\caption{\label{fig_arpesall_convergence}Averaged $A(\mathbf{k},\mathbf{k};E)$ for ensembles with different sample size or ensemble size, all generated with the same supercell size (SL10), filling factor ($n_c = 0.75$), and cluster size ($N_w = 4$): (a) $N_1 \times N_2 = 30 \times 30$, 20 samples; (b) $90 \times 90$, 20 samples; (c) $120 \times 120$, 20 samples; (d) $120 \times 120$, 50 samples. The color scale and path in $k$-space are the same as in the main text.}
\end{center}
\end{figure}

In Figs.~\ref{fig_arpesall_convergence}, we report a comparison of $A(\mathbf{k},\mathbf{k};E)$ plots obtained by varying the sample size or the ensemble size. For the case with $N_1 \times N_2 = 30 \times 30$ and 20 samples (Fig.~\ref{fig_arpesall_convergence}a), the number of $\mathbf{k}$ points is small and the dispersion looks quite vague (also note that this a zoomed view around the $K$ point, so that only about $1/4$ out of the $N_2$ points along the path are shown). Interestingly, the number of clusters seems to be already large enough to destroy the superlattice band structure: no repeated bands are visible, while the states are clearly arising from the graphene Dirac cone. The plot greatly improves when the size of the samples is increased to $90 \times 90$ (Fig.~\ref{fig_arpesall_convergence}b). However, further incresing the sample size (Fig.~\ref{fig_arpesall_convergence}c) or the number of samples (Fig.~\ref{fig_arpesall_convergence}d), gives only a slight improvement, which means that the result has already well converged.

\section{Procedure for band-gap extraction}

\begin{figure}[htp]
\begin{center}
\includegraphics[scale=0.59]{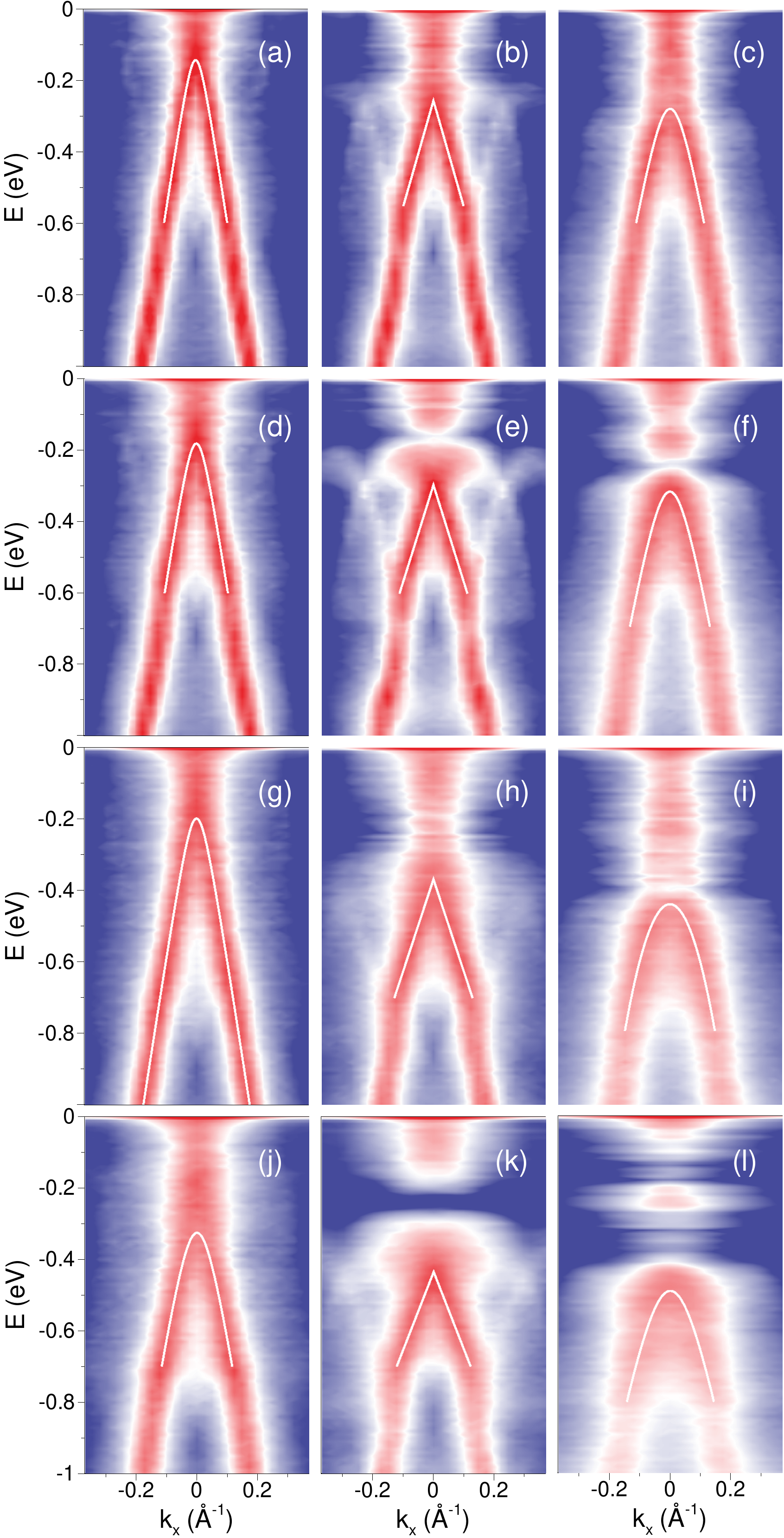}
\caption{\label{fig_arpesall}Plot of averaged $A(\mathbf{k},\mathbf{k};E)$ and fitting curve, for all the set of samples studied in this work. (i) and (l) are the same plots as in Fig.~1c of the manuscript, (j) and (k) the same as in Fig.~2a of the manuscript.}
\end{center}
\end{figure}

The band gap is extracted from each (ensemble-averaged) $A(\mathbf{k},\mathbf{k};E)$ plot using a fitting technique. We recall that the path in $k$-space is the one shown in the inset of Fig.~1c in the main text, so that $\mathbf{k} = (k_x, K_y)$, where $K_y$ is the $k_y$-coordinate of the $K$ point. Since $k_y$ is fixed, we use the simplified notation $A(k_x,k_x;E)$. The fitting procedure is composed of the following steps.
\begin{enumerate}
\item Manually choose a range of energies $[E_1,E_2]$ where to apply the fitting.
\item Find for each energy $E \in [E_1,E_2]$ the $k_x$ coordinate where the intensity is maximum, separately for positive and negative $k_x$:
\begin{eqnarray}
k_x^{+}(E) & \textrm{ such that } & A(k_x^{+},k_x^{+};E) = \max_{k_x \ge 0} A(k_x,k_x;E) \, , \label{eq_kp}\\
k_x^{-}(E) & \textrm{ such that } & A(k_x^{-},k_x^{-};E) = \max_{k_x \le 0} A(k_x,k_x;E) \, . \label{eq_km}
\end{eqnarray}
\item Compute for each $E \in [E_1,E_2]$ the values $w^+(E)$ and $w^-(E)$ as follows
\begin{align}
w^+(E) &= \frac{A\left(k_x^+(E),k_x^+(E);E\right)}{\max_{E' \in [E_1,E_2]} A\left(k_x^{+}(E),k_x^{+}(E);E'\right)} \, , \\
w^-(E) &= \frac{A\left(k_x^-(E),k_x^-(E);E\right)}{\max_{E' \in [E_1,E_2]} A\left(k_x^{-}(E),k_x^{-}(E);E'\right)} \, .
\end{align}
\item Apply a least-square fitting to the set of points $\left\{\left(E,k_x^+(E)\right)\right\} \cup \left\{\left(E,k_x^-(E)\right)\right\}$ with $E \in [E_1,E_2]$, by using $w^+(E)$ and $w^-(E)$ as weights and one of the following dispersion relations as fitting curve:
\begin{align}
E & = \pm \left( \hbar v |k_x| + \frac{E_g}{2} \right) \, , \label{eq_lin} \\
E & = \pm \left( \frac{\hbar^2 k_x^2}{2 m} + \frac{E_g}{2} \right) \, , \label{eq_par} \\
E & = \pm \sqrt{\frac{\hbar^2 E_g k_x^2}{2 m} + \left(\frac{E_g}{2}\right)^2} \, . \label{eq_lin_par} 
\end{align}
\end{enumerate}
The result of the fitting for negative $E$ is shown in Fig.~\ref{fig_arpesall}, superimposed to the original $A(\mathbf{k},\mathbf{k};E)$ plot.

For each $A(\mathbf{k},\mathbf{k};E)$ plot, a measure of the broadening is also extracted. We consider a specific $k_x$ value, $k_x^B$, and compute the quantity $2B$ as the difference between the two energies at which the function $A(k_x^B,k_x^B;E)$ decreases to half of its maximum value. 

The input and output parameters of the band-gap and broadening extraction are collected in Table~\ref{tab_extraction_parameters} for each set of sample: L, P, and LP refer to the fitting curves (\ref{eq_lin}), (\ref{eq_par}) and (\ref{eq_lin_par}), respectively, $v_F = (3/2) a_{\mathrm{CC}} |\gamma| / \hbar$ is the graphene Fermi velocity, and $m_0$ is the electron rest mass.
\begin{table}[tp]
\begin{center}
{
\begin{ruledtabular}
\begin{tabular}{cccccccccccc}
set & SL & $n_c$ & $N_w$ & $E_1$ & $E_2$ & fit. & $E_g/2$ & $v/v_F$ & $m/m0$ & $k_x^B$ & $2B$ \\
& & & & (eV) & (eV) & & (eV) & & & (\AA$^{-1}$) & (eV)\\
\hline
\hline
(a) & 13 & 0.75 & 2 & 0.05 & 0.6  & LP & 0.285 &       & 0.035 & 0.101 & 0.14 \\
(b) & 13 & 0.75 & 3 & 0.2  & 0.55 & L   & 0.507 & 0.537 &       & 0.101 & 0.15 \\
(c) & 13 & 0.75 & 4 & 0.25 & 0.6  & LP & 0.555 &       & 0.094 & 0.101 & 0.28 \\
(d) & 13 & 1    & 2 & 0.1  & 0.6  & LP & 0.365 &       & 0.046 & 0.101 & 0.13 \\
(e) & 13 & 1    & 3 & 0.3  & 0.6  & L   & 0.594 & 0.491 &       & 0.101 & 0.18 \\
(f) & 13 & 1    & 4 & 0.25 & 0.7  & LP & 0.633 &       & 0.110 & 0     & 0.12 \\
(g) & 10 & 0.75 & 2 & 0.1  & 1    & LP & 0.396 &       & 0.048 & 0.098 & 0.18 \\
(h) & 10 & 0.75 & 3 & 0.4  & 0.7  & L   & 0.736 & 0.465 &       & 0.098 & 0.18 \\
(i) & 10 & 0.75 & 4 & 0.4  & 0.8  & P   & 0.876 &       & 0.236 & 0     & 0.23 \\
(j) & 10 & 1    & 2 & 0.3  & 0.7  & LP & 0.650 &       & 0.087 & 0.098 & 0.22 \\
(k) & 10 & 1    & 3 & 0.35 & 0.7  & L   & 0.868 & 0.393 &       & 0.098 & 0.19 \\
(l) & 10 & 1    & 4 & 0.35 & 0.8  & LP & 0.979 &       & 0.192 & 0     & 0.23 
\end{tabular}
\end{ruledtabular}
}
\caption{\label{tab_extraction_parameters}}Parameters of the $A(\mathbf{k},\mathbf{k};E)$ fitting and broadening extraction.
\end{center}
\end{table}

The use of different fitting curves deserves an explanation. We notice that relation (\ref{eq_lin_par}) is the most physical one since it describes the 1D quantization of the graphene dispersion relation. However, for $N_w=3$, i.e. cases (b), (e), (h), and (k) of Fig.~\ref{fig_arpesall} and Table~\ref{tab_extraction_parameters}, the averaging effect of disorder at the cluster edges seems to be not strong enough to reach the typical dispersion relation (in fact, the repeated bands of the superlattice are still slightly visible in the $A(\mathbf{k},\mathbf{k};E)$ plot), and the functional dependence in (\ref{eq_lin}) gives a better fitting. Also, when then parabolicity is large in the energy range of interest, such as in cases (i) and (l) of Fig.~\ref{fig_arpesall} and Table~\ref{tab_extraction_parameters}, the use of (\ref{eq_par}) instead of (\ref{eq_lin_par}) does not make a significant difference in the gap value.

\section{Transport calculation}

\begin{figure}[b]
\begin{center}
\includegraphics[scale=0.75]{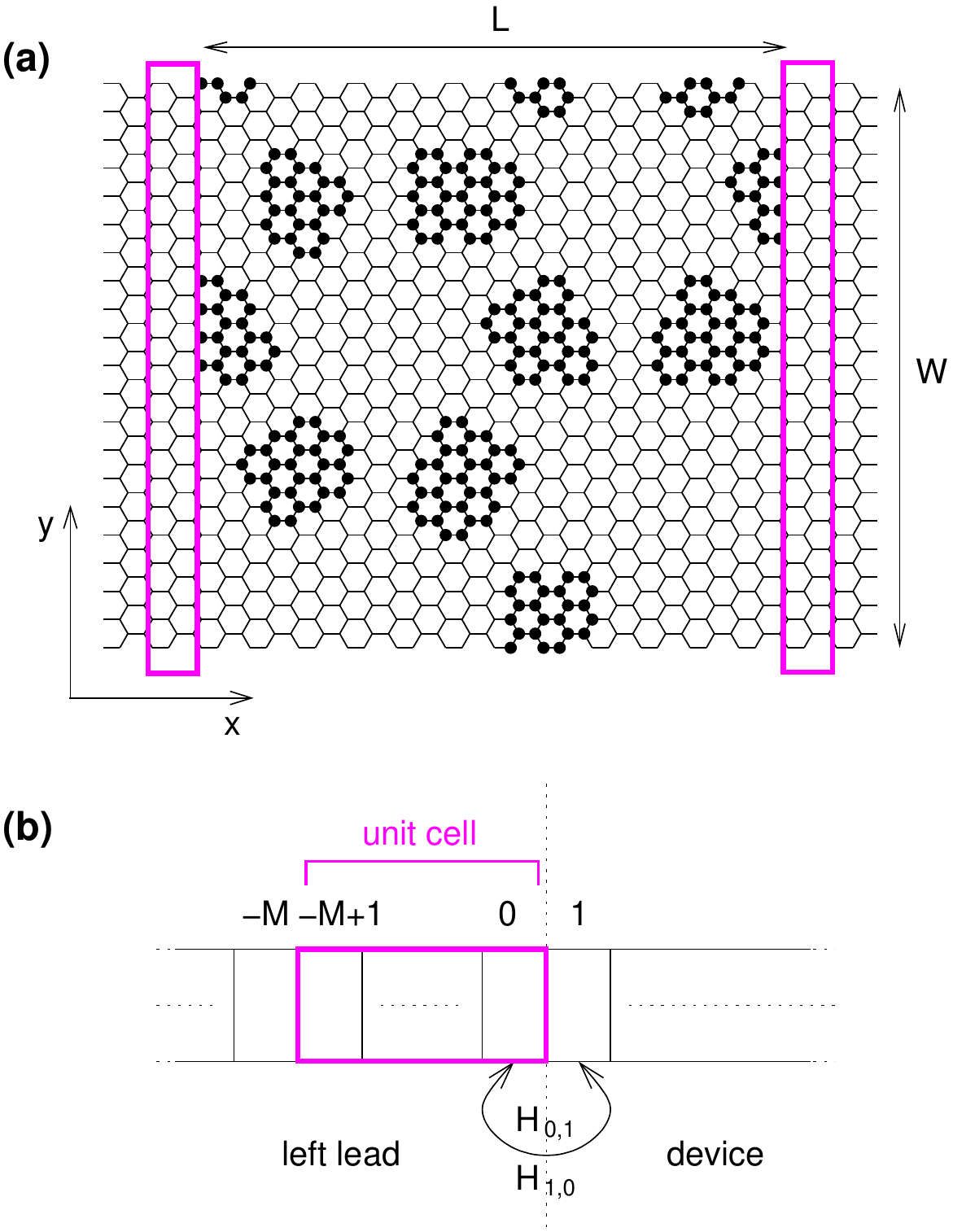}
\caption{(a) Example of device used in transport simulations: hydrogenated channel of size $W \times L$ between two leads of pristine graphene. The device is aligned so that its longitudinal direction corresponds to an armchair direction. (b) Slab representation of the same structure. The unit cell inside the lead regions can be viewed as being made of $M=4$ slabs, where each slab corresponds to a row of carbon atoms.} \label{fig_transport_structure}
\end{center}
\end{figure}

We consider a structure such as the one represented in Fig.~\ref{fig_transport_structure}a. The structure is divided in slabs along the longitudinal direction so that each slab corresponds to an atomic row. The slabs inside the device region are numbered from 1 to $N$.

The zero temperature conductance $G$ is given by the transmission function $T(E)$ at the Fermi energy $E_F$,
\begin{equation}
G = \frac{2e^2}{h} T(E = E_F) \, ,
\end{equation}
where in turn the transmission function is computed through Green's functions as \cite{datta97}
\begin{equation}
T(E) = \mathrm{Tr} \left[ \Gamma^L G^r \Gamma^R {G^a} \right] \, . \label{eq_transm1}
\end{equation}
In this equation, $\Gamma^{L/R} = \mathrm{i} (\Sigma^{r,L/R}-{\Sigma^{a,L/R}})$ is the broadening function due to the left/right lead, where $\Sigma^{r,L/R}$ is the self-energy representing the renormalization of the Hamiltonian of the device region alone due to the presence of the semi-infinite left/right lead, and $\Sigma^{a,L/R} = ( \Sigma^{r,L/R} )^\dagger$. Since the only non-null block of $\Sigma^{r,L}$ is $\Sigma^{r,L}_{1,1}$ and the only non-null one of $\Sigma^{r,R}$ is $\Sigma^{r,R}_{N,N}$, (\ref{eq_transm1}) can be rewritten as
\begin{eqnarray}
T(E) = \mathrm{Tr} \left[ \Gamma^L_{1,1} G^r_{1,N} \Gamma^R_{N,N} {G^a_{N,1}} \right] \, . \label{eq_transm2}
\end{eqnarray}
The calculation of the $G^r$ blocks can be efficiently performed using well known methods, such as the already mentioned recursive algorithm \cite{lake97}, or a combination of the recursive and the renormalization algorithms \cite{tony09PRB}, and therefore it is not treated here. Instead, we focus on the calculation of the lead self-energies.

We notice that in our case the unit cell of each lead is made of $M=4$ slabs  (Fig.~\ref{fig_transport_structure}b). Here, we propose an algorithm, based on the renormalization method \cite{grosso89}, to reduce the size of the unit cell to only one slab, such that the usual Sancho-Rubio algorithm \cite{sancho85} can then be applied on matrices having a reduced size, thus saving computational time. We notice that, in the specific case considered in this work, analytical espressions for the self-energies could have been used \cite{zhao09}. However, the numerical technique presented here is more general: for example, it can also be applied in the presence of a magnetic field.

We consider only the case of a left lead, the generalization to the right case being straightforward. The self-energy due to the left lead is defined as
\begin{eqnarray}
\Sigma^{r,L}_{1,1} = \Omega_{1,0} g^r_{0,0} \Omega_{0,1} \, ,
\end{eqnarray}
where $\Omega = (E+\mathrm{i}\eta)I-H$ and $g^r$ is the retarded Green's function for the case in which the coupling between the device and the leads is set to zero \cite{datta97}. Suppose that the unit cell of the lead contains $M$ slabs. The matrix $\Omega^L$ of the isolated left lead has thus the structure
\begin{eqnarray}
\Omega^L = \left( \begin{array}{cccc}
\ddots & \ddots      &             &             \\
\ddots & \ddots      & \Omega_{-M,-M+1} &             \\
       & \Omega_{-M+1,-M} & \Omega^C         & \Omega_{-M,-M+1} \\
       &             & \Omega_{-M+1,-M} & \Omega^C
\end{array} \right) \, ,
\end{eqnarray}
with
\begin{eqnarray}
\Omega^C = \left( \begin{array}{cccc}
\Omega_{-M+1,-M+1} & \Omega_{-M+1,-M+2} &           &          \\
\Omega_{-M+2,-M+1} & \ddots        & \ddots    &          \\
              & \ddots        & \Omega_{-1,-1} & \Omega_{-1,0} \\
              &               & \Omega_{0,-1}  & \Omega_{0,0}
\end{array} \right) \, .
\end{eqnarray}
As a first step, we consider $\Omega^C$ and decimate all the slabs from $-1$ backward to $-M+2$ (assuming $M > 2$). We define $d_1^{(0)} = \Omega_{0,0}$, $d_2^{(0)} = \Omega_{-1,-1}$, $a^{(0)} = \Omega_{-1,0}$, $b^{(0)} = \Omega_{0,-1}$. The generic iteration of index $n$ ($n = 1, \ldots, M-2$) consists in eliminating the second last node from the matrix
\begin{eqnarray}
\left( \begin{array}{cccc}
\ddots & \ddots        &             &             \\
\ddots & \Omega_{-n-1,-n-1} & \Omega_{-n-1,-n} &             \\
       & \Omega_{-n,-n-1}   & d_2^{(n-1)} & a^{(n-1)}   \\
       &               & b^{(n-1)}   & d_1^{(n-1)} \\
\end{array} \right)
\end{eqnarray}
with the equations
\begin{eqnarray}
d_1^{(n)} \! &=& \! d_1^{(n-1)} - b^{(n-1)} \left( d_2^{(n-1)} \right)^{-1} a^{(n-1)} \, , \nonumber \\
d_2^{(n)} \! &=& \! \Omega_{-n-1,-n-1} - \, \Omega_{-n-1,-n} \left( d_2^{(n-1)} \right)^{-1} \Omega_{-n,-n-1} \, , \nonumber \\
a^{(n)} \! &=& \! -\Omega_{-n-1,-n} \left( d_2^{(n-1)} \right)^{-1} a^{(n-1)} \, , \nonumber \\
b^{(n)} \! &=& \! -b^{(n-1)} \left( d_2^{(n-1)} \right)^{-1} \Omega_{-n,-n-1} \, ,
\end{eqnarray}
which are simply an application of (\ref{eq_decimation}).
At the end, we obtain the renormalized $\Omega^L$ matrix
\begin{eqnarray}
\tilde{\Omega}^L = \left( \begin{array}{ccccc}
\ddots & \ddots      &             &             &             \\
\ddots & d_2^{(M-2)} & a^{(M-2)}   &             &             \\
       & b^{(M-2)}   & d_1^{(M-2)} & \Omega_{-M,-M+1} &             \\
       &             & \Omega_{-M+1,-M} & d_2^{(M-2)} & a^{(M-2)}   \\
       &             &             & b^{(M-2)}   & d_1^{(M-2)} \\
\end{array} \right) \, .
\end{eqnarray}
As a second step, we consider $\tilde{\Omega}^L$ and decimate all the even slabs (assuming $M > 1$). By using the formulas (again an application of Eqs.~\ref{eq_decimation})
\begin{eqnarray}
\delta_1^{(0)} \! &=& \! d_1^{(M-2)} - b^{(M-2)} \left( d_2^{(M-2)} \right)^{-1} a^{(M-2)} \, , \nonumber \\
\delta_2^{(0)} \! &=& \! \delta_1^{(0)} - \Omega_{-M,-M+1} \left( d_2^{(M-2)} \right)^{-1} \Omega_{-M+1,-M} \, , \nonumber \\
\alpha^{(0)} \! &=& \! -\Omega_{-M,-M+1} \left( d_2^{(M-2)} \right)^{-1} a^{(M-2)} \, , \nonumber \\
\beta^{(0)} \! &=& \! -b^{(M-2)} \left( d_2^{(M-2)} \right)^{-1} \Omega_{-M+1,-M} \, ,
\end{eqnarray}
we get a new renormalized $\Omega^L$ matrix,
\begin{eqnarray}
\tilde{\tilde{\Omega}}^L = \left( \begin{array}{ccccc}
\ddots & \ddots         &                &                &                \\
\ddots & \delta_2^{(0)} & \alpha^{(0)}   &                &                \\
       & \beta^{(0)}    & \delta_2^{(0)} & \alpha^{(0)}   &                \\
       &                & \beta^{(0)}    & \delta_2^{(0)} & \alpha^{(0)}   \\
       &                &                & \beta^{(0)}    & \delta_1^{(0)} \\
\end{array} \right) \, .
\end{eqnarray}
This matrix has the same structure as the one used in the Sancho-Rubio algorithm \cite{sancho85}. The generic iteration of index $n$ ($n = 1,2,\ldots$) of this algorithm actually consists in the decimation of the slabs with even indexes from the matrix
\begin{eqnarray}
\left( \begin{array}{ccccc}
\ddots & \ddots           &                  &                  &                  \\
\ddots & \delta_2^{(n-1)} & \alpha^{(n-1)}   &                  &                  \\
       & \beta^{(n-1)}    & \delta_2^{(n-1)} & \alpha^{(n-1)}   &                  \\
       &                  & \beta^{(n-1)}    & \delta_2^{(n-1)} & \alpha^{(n-1)}   \\
       &                  &                  & \beta^{(n-1)}    & \delta_1^{(n-1)} \\
\end{array} \right) \, ,
\end{eqnarray}
by using the formulas (again from Eqs.~\ref{eq_decimation})
\begin{eqnarray}
\delta_1^{(n)} \! &=& \! \delta_1^{(n-1)} - \beta^{(n-1)} \left( \delta_2^{(n-1)} \right)^{-1} \alpha^{(n-1)} \, , \nonumber \\
\delta_2^{(n)} \! &=& \! \delta_2^{(n-1)} - \beta^{(n-1)} \left( \delta_2^{(n-1)} \right)^{-1} \alpha^{(n-1)} - \, \alpha^{(n-1)} \left( \delta_2^{(n-1)} \right)^{-1} \beta^{(n-1)}                  \, , \nonumber \\
\alpha^{(n)}   \! &=& \! - \alpha^{(n-1)} \left( \delta_2^{(n-1)} \right)^{-1} \alpha^{(n-1)}                 \, , \nonumber \\
\beta^{(n)}    \! &=& \! - \beta^{(n-1)} \left( \delta_2^{(n-1)} \right)^{-1} \beta^{(n-1)}                   \, ,
\end{eqnarray}
until convergence, i.e. until the coupling matrices $\alpha^{(n)}$ and $\beta^{(n)}$ become sufficiently small. At the end, we can approximate $g^r_{0,0} = \left( \delta_1^{(n)} \right)^{-1}$, where $n$ stands for the index of the last iteration.

\providecommand{\noopsort}[1]{}\providecommand{\singleletter}[1]{#1}%

\end{document}